\documentclass[aps,showpacs,twocolumn]{revtex4}
\usepackage{graphicx}

\begin{document}

\title{ Can accretion disk properties distinguish gravastars from black
holes?}
\author{Tiberiu Harko}
\email{harko@hkucc.hku.hk}
\affiliation{Department of Physics and Center for Theoretical
and Computational Physics,
The University of Hong Kong, Pok Fu Lam Road, Hong Kong}
\author{Zolt\'{a}n Kov\'{a}cs}
\email{zkovacs@mpifr-bonn.mpg.de}
\affiliation{Max-Planck-Institute f\"{u}r Radioastronomie,
Auf dem H\"{u}gel 69, 53121
Bonn, Germany}
\affiliation{Department of Experimental Physics, University of Szeged,
D\'{o}m T\'{e}r 9,
Szeged 6720, Hungary}
\author{Francisco S. N. Lobo}
\email{flobo@cii.fc.ul.pt} \affiliation{Centro de F\'{i}sica
Te\'{o}rica e Computacional, Faculdade de Ci\^{e}ncias da
Universidade de Lisboa, Avenida Professor Gama Pinto 2, P-1649-003
Lisboa, Portugal}
\date{\today}

\begin{abstract}

Gravastars, hypothetic astrophysical objects, consisting of a dark
energy condensate surrounded by a strongly correlated thin shell
of anisotropic matter, have been proposed as an alternative to the
standard black hole picture of general relativity. Observationally
distinguishing between astrophysical black holes and gravastars is
a major challenge for this latter theoretical model. This due to
the fact that in static gravastars large stability regions (of the
transition layer of these configurations) exist that are
sufficiently close to the expected position of the event horizon,
so that it would be difficult to distinguish the exterior geometry
of gravastars from an astrophysical black hole. However, in the
context of stationary and axially symmetrical geometries, a
possibility of distinguishing gravastars from black holes is
through the comparative study of thin accretion disks around
rotating gravastars and Kerr-type black holes, respectively. In
the present paper, we consider accretion disks around slowly
rotating gravastars, with all the metric tensor components
estimated up to the second order in the angular velocity. Due to
the differences in the exterior geometry, the thermodynamic and
electromagnetic properties of the disks (energy flux, temperature
distribution and equilibrium radiation spectrum) are different for
these two classes of compact objects, consequently giving clear
observational signatures. In addition to this, it is also shown
that the conversion efficiency of the accreting mass into
radiation is always smaller than the conversion efficiency for
black holes, i.e., gravastars provide a less efficient mechanism
for converting mass to radiation than black holes. Thus, these
observational signatures provide the possibility of clearly
distinguishing rotating gravastars from Kerr-type black holes.

\end{abstract}

\maketitle


\section{Introduction}

 The Schwarzschild solution has played a fundamental conceptual role in
general relativity, and beyond, for instance, regarding event
horizons, spacetime singularities, and aspects of quantum field
theory in curved spacetimes. However, one still encounters in the
literature the existence of misconceptions, as well as a certain
ambiguity inherent in the Schwarzschild solution (we refer the
reader to \cite{Doran} for a detailed review). In this context,
recently a new final state of gravitational collapse has been
proposed, denoted as {\it gravastars} ({\it grav}itational {\it
va}cuum {\it stars}) \cite{MaMo04}, which represent a viable
alternative to black holes, and their properties have been
extensively investigated. These models consist of a compact object
with an interior de Sitter condensate, governed by an equation of
state given by $p=-\rho$, matched to a shell of finite thickness
with an equation of state $p=\rho$. The latter is then matched to
an exterior Schwarzschild vacuum solution. The thick shell
replaces both the de Sitter and the Schwarzschild horizons,
therefore, the gravastar model has no singularity at the origin
and no event horizon, as its rigid surface is located at a radius
slightly greater than the Schwarzschild radius \cite{MaMo04}.
These configurations are stable from a thermodynamic point of
view. The issue of the dynamic stability of the transition layer
(an infinitesimally thin shell) against spherically symmetric
perturbations was considered in \cite{ViWi04}, by constructing a
model that shares the key features of the gravastar scenario. It
was found that there are some physically reasonable equations of
state for the transition layer that lead to stability. This latter
stability analysis was further generalized to an anti-de Sitter or
de Sitter interior and a Schwarzschild (anti)-de Sitter or
Reissner-Nordstr\"{o}m exterior \cite{Ca05}. Recently, dynamical
models of prototype gravastars were constructed and studied
\cite{Rocha:2008hi}. It was found that in some cases the models
represent stable gravastars, while in other cases they represent
``bounded excursion'' stable gravastars, where the thin shell is
oscillating between two finite radii. In some other cases they
collapse until the formation of black holes.

In addition to this, gravastar models that exhibit continuous
pressure without the presence of infinitesimally thin shells were
introduced in \cite{CaFaVi05} and further analyzed in
\cite{DeBenedictis:2005vp}. By considering the usual TOV equation
for static solutions with negative central pressure, it was found
that gravastars cannot be perfect fluids and anisotropic pressures
in the `crust' of a gravastar-like object are unavoidable. The
anisotropic TOV equation can then be used to bound the pressure
anisotropy, and the transverse stresses that support a gravastar
permit a higher compactness than the Buchdahl-Bondi bound for
perfect-fluid stars. A wide variety of gravastar models within the
context of nonlinear electrodynamics were also constructed in
\cite{LoAr07}. Using the $F$ representation, specific forms of
Lagrangians were considered describing magnetic gravastars, which
may be interpreted as self-gravitating magnetic monopoles with
charge $g$. Using the dual $P$ formulation of nonlinear
electrodynamics, electric gravastar models were constructed by
considering specific structural functions, and the characteristics
and physical properties of the solutions were further explored.
Gravastar solutions with a Born-Infeld phantom replacing the de
Sitter interior were also analyzed in \cite{Bilic:2005sn}.

It has also been recently proposed by Chapline that this new
emerging picture consisting of a compact object resembling
ordinary spacetime, in which the vacuum energy is much larger than
the cosmological vacuum energy, has been denoted as a `dark energy
star' \cite{Chapline}. Indeed, a generalization of the gravastar
picture was considered in \cite{Lo06} by considering a matching of
an interior solution governed by the dark energy equation of
state, $\omega =p/\rho < -1/3$, to an exterior Schwarzschild
vacuum solution at a junction interface. Several relativistic dark
energy stellar configurations were analyzed by imposing specific
choices for the mass function, by assuming a constant energy
density, and a monotonic decreasing energy density in the star's
interior, respectively. The dynamical stability of the transition
layer of these dark energy stars to linearized spherically
symmetric radial perturbations about static equilibrium solutions
was also considered, and it was found that large stability regions
exist that are sufficiently close to where the event horizon is
expected to form. Evolving dark energy stars were explored in
\cite{DeBenedictis:2008qm}, where a time-dependent dark energy
parameter was considered. The general properties of a spherically
symmetric body described through the generalized Chaplygin
equation of state were also extensively analyzed in
\cite{Bertolami:2005pz}. In the context of cosmological equations
of state, in \cite{Lobo:2006ue} the construction of gravastars
supported by a van der Waals equation of state was studied and
their respective characteristics and physical properties were
further analyzed. It was argued that these {\it van der Waals
quintessence stars} may possibly originate from density
fluctuations in the cosmological background. Note that the van der
Waals quintessence equation of state is an interesting scenario
that describes the late universe, and seems to provide a solution
to the puzzle of dark energy, without the presence of exotic
fluids or modifications of the Friedmann equations.

However, observationally distinguishing between astrophysical
black holes and gravastars is a major challenge for this latter
theoretical model, as in static gravastars large stability regions
exist that are sufficiently close to the expected position of the
event horizon. The constraints that present-day observations of
well-known black hole candidates place on the gravastar model were
discussed in \cite{BrNa07}. The heating of neutron stars via
accretion is well documented by astronomical observations. A
gravastar would acquire most of its mass via accretion, either as
part of its birth (e.g., during core collapse in a supernova
explosion followed by the rapid accretion of a fallback disk), or
over an extended period of time after birth. Similar heating
processes have not been observed in the case of black hole
candidates. However, the absence of a detectable heating may also
be consistent with a gravastar, if the heat capacity is large
enough that it requires large amounts of heat to produce small
changes in temperature. Nevertheless, some level of accretion
heating is unavoidable, and provides some strong constraints on
the gravastar model. The large surface redshifts employed in
gravastar models implies that the internal energy generated per
unit rest mass accreted is very nearly $c^{2}$, and that the
radiation emitted from the surface of the object should be an
almost perfect black body. The energy evolution of an accreting
gravastar is determined by the equation $dU/dt=\dot{M}c^{2}-L$,
where $U$ is the internal energy, $\dot{M}$ is the mass accretion
rate, and $L$ is the luminosity. Assuming that the gravastar is
the result of a BEC-like phase transition induced by strong
gravity \cite{Chapline}, the gravastar which starts at zero
temperature and rapidly accretes a mass $\Delta mM_{Sun}$ will be
heated to a temperature as observed at infinity of $T_{h}\approx
3.1\times 10^{6}m^{-2/3}\xi ^{1/3}\left( \Delta m/m\right) ^{1/3}$
K, where $m$ is the mass of the gravastar in solar mass units, and
$\xi =l/l_{Pl}$ is the length scale in Planck units at which
general relativity fails to adequately describe gravity. In the
case of slow accretion, the temperature is given by the
equilibrium value $T_{eq}\approx 1.7\times 10^{7}\left(
\dot{M}/\dot{M} _{Edd}\right) ^{1/4}m^{-1/4}$ K, where
$\dot{M}_{Edd}=2.3\times10^{-9}mM_{\odot}/{\rm yr}$ is the
Eddington mass accretion rate. By comparing these theoretical
predictions on the temperature with the upper limits of the
temperature of the black hole candidates, obtained through
observations, one can find some limits on $ \xi $. Two black hole
candidates, known to have extraordinarily low luminosities,
namely, the supermassive black hole in the galactic center,
Sagittarius A*, and the stellar-mass black hole, XTE J1118 + 480,
respectively, were considered in the analysis \cite{BrNa07}. For
XTE J1118+480, due to the low mass and low heat capacity, the
value of $\xi $ is constrained to $\xi \approx 5\times 10^3$,
while for Sagittarius A* values for $\xi $ in the range $10^4$ and
$10^{11}$ are excluded. Therefore the length scale for the
considered gravastar models must be sub-Planckian. Thus, if any
significant fraction of the mass of the gravastars is due to
accretion, we should see the thermal emission associated with that
phase, which is ruled out for Sagittarius A* and XTE J1118 + 480.

The question of whether gravastars can be distinguished from black
holes at all was also considered, from a theoretical point of
view, in \cite{ChRe07}, where two basic questions analyzed were:
(i) Is a gravastar stable against generic perturbations and, (ii)
if it is stable, can an observer distinguish it from a black hole
of the same mass? A general class of gravastars was constructed,
and the equilibrium conditions in order to exist as solutions of
the Einstein equations were obtained. It was found that gravastars
are stable to axial perturbations, and that their quasi-normal
modes differ from those of a black hole of the same mass. Thus,
these modes can be used to discern, beyond dispute, a gravastar
from a black hole. The formation hysteresis effects were ignored
in this study. In addition to this, sharp analytic bounds on the
surface compactness $2m/r$ that follow from the requirement that
the dominant energy condition (DEC) holds at the shell were
derived in \cite{HoIl07}. In the case of a Schwarzschild exterior,
the highest surface compactness is achieved with the stiff shell
in the limit of vanishing (dark) energy density in the interior.
In the case of a Schwarzschild-de Sitter exterior, it was shown
that gravastar configurations with a surface pressure and with a
vanishing shell pressure (dust shells), are allowed by the DEC.
The causality requirement (sound speed not exceeding that of
light) further restricts the space of allowed gravastar
configurations.

The ergoregion instability is known to affect very compact objects
that rotate very rapidly, and that do not possess an horizon. A
detailed analysis on the relevance of the ergoregion instability
for the viability of gravastars was presented in
\cite{Cardoso:2007az,ChRe08}. In \cite{Cardoso:2007az}, it was
shown that ultra-compact objects with high redshift at their
surface are unstable when rapidly spinning, which strengthens the
role of black holes as candidates for astrophysical observations
of rapidly spinning compact objects. In particular, analytical and
numerical results indicate that gravastars are unstable against
scalar field perturbations. Their instability timescale is many
orders of magnitude stronger than the instability timescale for
ordinary stars with uniform density. In the large $l = m$
approximation, suitable for a WKB treatment, gravitational and
scalar perturbations have similar instability timescales. In the
low-$m$ regime gravitational perturbations are expected to have
shorter instability timescales than scalar perturbations.
In \cite{ChRe08}, the analysis shows that not all rotating
gravastars are unstable, and stable models can be constructed also
with $ J/M^2\sim 1$, where $J$ and $M$ are the angular momentum
and mass of the gravastar, respectively. Therefore, the existence
of gravastars cannot be ruled out by invoking the ergoregion
instability. The gravastar model was extended by introducing an
electrically charged component in \cite{HoIlMa09}, where the
Einstein--Maxwell field equations were solved in the
asymptotically de Sitter interior, and a source of the electric
field was coupled to the fluid energy density. Two different
solutions that satisfy the dominant energy condition were given,
and the equation of state, the speed of sound and the surface
redshift were calculated for both models. The dipolar magnetic
field configuration for gravastars was studied in \cite{TuAhAb09},
and solutions of Maxwell equations in the internal background
spacetime of a slowly rotating gravastar were obtained. The shell
of the gravastar where the magnetic field penetrated was modeled
as a sphere consisting of a highly magnetized perfect fluid, with
infinite conductivity. It was assumed that the dipolar magnetic
field of the gravastar is produced by a circular current loop
symmetrically placed at radius $a$ at the equatorial plane.

The mass accretion around rotating black holes was studied in
general relativity for the first time in \cite{NoTh73}. By using
an equatorial approximation to the stationary and axisymmetric
spacetime of rotating black holes, steady-state thin disk models
were constructed, extending the theory of non-relativistic
accretion \cite{ShSu73}. In these models hydrodynamical
equilibrium is maintained by efficient cooling mechanisms via
radiation transport, and the accreting matter has a Keplerian
rotation. The radiation emitted by the disk surface was also
studied under the assumption that black body radiation would
emerge from the disk in thermodynamical equilibrium. The radiation
properties of thin accretion disks were further analyzed  in
\cite{PaTh74,Th74}, where the effects of photon capture by the
hole on the spin evolution were presented as well. In these works
the efficiency with which black holes convert rest mass into
outgoing radiation in the accretion process was also computed.
More recently, the emissivity properties of the accretion disks
were investigated for exotic central objects, such as wormholes
\cite{Harko:2008vy}, and non-rotating or rotating quark, boson or
fermion stars and brane-world black holes
\cite{Bom,To02,YuNaRe04,Guzman:2005bs,Pun:2008ua,KoChHa09}. The
radiation power per unit area, the temperature of the disk and the
spectrum of the emitted radiation were given, and compared with
the case of a Schwarzschild black hole of an equal mass. The
physical properties of matter forming a thin accretion disk in the
static and spherically symmetric spacetime metric of vacuum $f(R)$
modified gravity models were also analyzed \cite{Pun:2008ae}, and
it was shown that particular signatures can appear in the
electromagnetic spectrum, thus leading to the possibility of
directly testing modified gravity models by using astrophysical
observations of the emission spectra from accretion disks.

It is the purpose of the present paper to consider another
observational possibility that may distinguish gravastars from
black holes, namely, the study of the properties of the thin
accretion disks around rotating gravastars and black holes,
respectively. Thus, we consider a comparative study of the thin
accretion disks around slowly rotating gravastars and black holes,
respectively. In particular, we consider the basic physical
parameters describing the disks, such as the emitted energy flux,
the temperature distribution on the surface of the disk, as well
as the spectrum of the emitted equilibrium radiation. Due to the
differences in the exterior geometry, the thermodynamic and
electromagnetic properties of the disks (energy flux, temperature
distribution and equilibrium radiation spectrum) are different for
these two classes of compact objects, thus giving clear
observational signatures, which may allow to discriminate, at
least in principle, gravastars from black holes. We would like to
point out that the proposed method for the detection of the
gravastars by studying accretion disks is an {\it indirect}
method, which must be complemented by {\it direct} methods of
observation of the surface of the considered compact objects.

The present paper is organized as follows. In Sec. \ref{sec:II},
we present the fundamental field equations for static and slowly
rotating gravastar models. In Sec. \ref{sec:III}, we review the
formalism and the physical properties of the thin disk accretion
onto compact objects, for stationary axisymmetric spacetimes. In
Sec.~\ref{sec:IV}, we analyze the basic properties of matter
forming a thin accretion disk around slowly rotating gravastar
spacetimes. We discuss and conclude our results in Sec.
\ref{sec:concl}. Throughout this work, we use a system of units so
that $c=G=\hbar =k_{B}=1$, where $k_{B}$ is Boltzmann's constant.

\section{Slowly rotating gravastar and Kerr black holes}
\label{sec:II}

In order to construct slowly rotating gravastar models we first
consider the static case. Then, by assuming that rotation
represents a second order perturbation of the static case, a
slowly rotating gravastar model can be constructed.

\subsection{Static gravastar models}

\subsubsection{Equations of
structure}\label{sec:graveqs}

For a static general relativistic spherically symmetric matter
configuration, the interior line element can be taken generally as
\begin{equation}
ds^{2}=-e^{\nu (r)}dt^{2}+e^{\lambda (r)}dr^{2}+r^{2}\left( d\theta
^{2}+\sin ^{2}\theta d\varphi ^{2}\right) .
\end{equation}
We assume that the star consists of an anisotropic fluid
distribution of matter, and is given by
\begin{equation}
T_{\mu\nu}=(\rho+p_{\perp})U_\mu \, U_\nu+p_{\perp}\,
g_{\mu\nu}+(p_r-p_{\perp})\chi_\mu \chi_\nu \,,
\end{equation}
where $U^\mu$ is the four-velocity, $\chi^\mu$ is the unit
spacelike vector in the radial direction, i.e.,
$\chi^\mu=e^{-\lambda (r)/2}\,\delta^\mu{}_r$. $\rho(r)$ is the
energy density, $p_r(r)$ is the radial pressure measured in the
direction of $\chi^\mu$, and $p_{\perp}(r)$ is the transverse
pressure measured in the orthogonal direction to $\chi^\mu$.
Taking into account the above considerations, the stress-energy
tensor is given by the following profile: $T^{\mu}{}_{\nu}={\rm
diag}[-\rho(r),p_r(r),p_{\perp}(r),p_{\perp}(r)]$.

We suppose that inside the star $p_{r}\neq p_{\perp }$, $\forall
r\neq 0$, and define the anisotropy parameter as $\Delta =p_{\perp
}-p_{r}$, where $\Delta $ is a measure of the deviations from
isotropy. If $\Delta >0,\forall r\neq 0$ the body is tangential
pressure dominated while $\Delta <0$ indicates that $
p_{r}>p_{\perp }$. Note that $\Delta/r$ represents a force due to
the anisotropic nature of the stellar model, which is repulsive,
i.e., being outward directed if $p_{\perp}>p_r$, and attractive if
$p_{\perp}<p_r$.

The properties of the anisotropic compact object can be completely
described by the gravitational structure equations, which are
given by:
\begin{equation}  \label{5}
\frac{dm}{dr}=4\pi \rho r^{2},
\end{equation}
\begin{equation}  \label{6}
\frac{dp_{r}}{dr}=-\frac{\left (\rho +p_{r}\right )\left [m+4\pi r^{3}
p_{r}\right ]}{r^{2}\left (1-\frac{2m}{r}\right )}+\frac{2\Delta }{r},
\end{equation}
\begin{equation}  \label{7}
\frac{d\nu }{dr}=-\frac{2}{\rho
+p_{r}}\frac{dp_{r}}{dr}+\frac{4\Delta }{ r\left (\rho
+p_{r}\right )},
\end{equation}
where $m(r)$ is the mass inside radius $r$, and the relationship
$e^{-\lambda(r)}=[1-2m(r)/r]$ has been used.

A solution of Eqs.~(\ref{5})-(\ref{7}) is possible only when
boundary conditions have been imposed. As in the isotropic case we
require that the interior of any matter distribution be free of
singularities, which imposes the condition $m(r)\rightarrow 0$ as
$r\rightarrow 0$. Assuming that $p_{r}$ is finite at $r=0$, we
have $\nu ^{\prime }\rightarrow 0$ as $r\rightarrow 0$. Therefore
the gradient $dp_{r}/dr$ will be finite at $r=0$ only if $\Delta $
vanishes at least as rapidly as $r$ when $r\rightarrow 0$. This
requires that the anisotropy parameter satisfies the boundary
condition
\begin{equation}
\lim _{r\rightarrow 0}\frac{\Delta \left (r\right )}{r}=0.
\end{equation}

At the center of the star the other boundary conditions for
Eqs.~(\ref{5})-(\ref{7}) are $p_{r}(0)=p_{\perp }(0)=p_{c}$ and
$\rho (0)=\rho _{c}$, where $ \rho _{c}$ and $p_{c}$ are the
central density and pressure, respectively. The radius $a$ of the
star is determined by the boundary condition $ p_{r}\left(
a\right) =0$. We do not necessarily require that the tangential
pressure $p_{\perp }$ vanishes for $r=a$. Therefore at the surface
of the star the anisotropy parameter satisfies the boundary
condition $\Delta (a)=p_{\perp }(a)-p_{r}(a)=p_{\perp }(a)\geq 0$.
To close the field equations the equations of state of the radial
pressure $p_{r}=p_{r}\left( \rho \right) $ and of the tangential
pressure $p_{\perp}=p_{\perp }\left( \rho \right) $ must also be
given. To be a gravastar model, we need to impose the equation of
state $p_r=-\rho$, so that from the field equations one easily
deduces that
\begin{equation}
\nu(r)=-\lambda(r)=\ln\left[1-\frac{2m(r)}{r}\right]\,,
\end{equation}
where $m(r)$ is the mass function.

\subsubsection{Junction interface}

We consider models of gravastars by matching an interior solution,
governed by an equation of state, $p_r=-\rho$, to an exterior
Schwarzschild vacuum solution with $p=\rho=0$, at a junction
interface $\Sigma$, with junction radius $a$. The Schwarzschild
metric is given by
\begin{equation}
ds^2=-\left(1-\frac{2M}{r}\right)\,dt^2+\left(1-
\frac{2M}{r}\right)^{-1}dr^2+r^2 \,d\Omega^2 \label{Sch-metric}\,,
\end{equation}
which possesses an event horizon at $r_b=2M$, and
$d\Omega^2=d\theta ^2+\sin ^2{\theta} \, d\varphi ^2$. To avoid
the event horizon, the junction radius lies outside $2M$, i.e.,
$a>2M$. We show below that $M$, in this context, may be
interpreted as the total mass of the gravastar.

Using the Darmois-Israel formalism~\cite{Darmois-Israel}, the
surface stresses of the thin shell are given by \cite{Lo06}
\begin{eqnarray}
&\sigma=-\frac{1}{4\pi a} \left(\sqrt{1-\frac{2M}{a}+\dot{a}^2}-
\sqrt{1-\frac{2m}{a}+\dot{a}^2} \, \right)
    \label{surfenergy}   ,\\
&{\cal P}=\frac{1}{8\pi a} \Bigg(\frac{1-\frac{M}{a}
+\dot{a}^2+a\ddot{a}}{\sqrt{1-\frac{2M}{a}+\dot{a}^2}}
      - \frac{1-m'-\frac{m}{a}+\dot{a}^2+a\ddot{a}}
{\sqrt{1-\frac{2m}{a}+\dot{a}^2}} \, \Bigg)
    \label{surfpressure}    \,,
\end{eqnarray}
where the overdot denotes a derivative with respect to proper
time, $\tau$. $\sigma$ and ${\cal P}$ are the surface energy
density and the tangential pressure, respectively~\cite{Lo06}. The
dynamical stability of the transition layer of these compact
spheres to linearized spherically symmetric radial perturbations
about static equilibrium solutions was explored using Eqs.
(\ref{surfenergy})-(\ref{surfpressure}) (see ~\cite{Lo06} for
details). Large stability regions were found that exist
sufficiently close to where the event horizon is expected to form,
so that it would be difficult to distinguish the exterior geometry
of these gravastars from astrophysical black holes.

The surface mass of the thin shell is given by $m_s=4\pi a^2
\sigma$. By rearranging Eq. (\ref{surfenergy}), evaluated at a
static solution $a_0$, i.e., $\dot{a}=\ddot{a}=0$, one obtains the
total mass of the gravastar, given by
\begin{equation}\label{totalmass}
M=m(a_0)+m_s(a_0)\left[\sqrt{1-\frac{2m(a_0)}{a_0}}
-\frac{m_s(a_0)}{2a_0}\right]
\,.
\end{equation}

\subsubsection{Specific model: Tolman-Matese-Whitman mass
function}

To gain insight into the problem, it is interesting to present a
specific example. For instance, consider the following choice for
the mass function, given by
\begin{equation}
m(r)=\frac{b_0 r^3}{2(1+2b_0 r^2)}\,,
      \label{TMWmass}
\end{equation}
where $b_0$ is a non-negative constant, which was extensively
analyzed in \cite{Lo06} in the context of dark energy stars. The
latter may be determined from the regularity conditions and the
finite character of the energy density at the origin $r=0$, and is
given by $b_0=8\pi \rho_c/3$, where $\rho_c$ is the energy density
at $r=0$.

This choice of the mass function represents a monotonic decreasing
energy density in the star interior, and was used previously in
the analysis of isotropic fluid spheres by Matese and Whitman
\cite{MatWhit} as a specific case of the Tolman type$-IV$ solution
\cite{Tolman}, and later by Finch and Skea~\cite{Finch}.
Anisotropic stellar models, with the respective astrophysical
applications, were also extensively analyzed in Refs. \cite{Mak},
by considering a specific case of the Matese-Whitman mass
function. The numerical results outlined show that the basic
physical parameters, such as the mass and radius, of the model can
describe realistic astrophysical objects such as neutron stars
\cite{Mak}.

The spacetime metric for this solution is provided by
\begin{eqnarray}
&ds^2=-\left(\frac{1+b_0r^2}{1+2b_0r^2}\right)\,dt^2
    +\left(\frac{1+2b_0r^2}{1+b_0r^2}\right)dr^2+r^2
\,d\Omega^2   \,.
\end{eqnarray}
The stress-energy tensor components are given by
\begin{eqnarray}
p_r&=&-\rho =-\left( \frac{b_0}{8\pi}\right)\frac{\left(3+2b_0
r^2\right)}{(1+2b_0 r^2)^2}
    \nonumber   \\
p_{\perp}&=&-\left( \frac{b_0}{8\pi}\right)\frac{(3+2b_0
r^2)}{(1+2b_0 r^2)^2}+\left(\frac{b_0^2r^2}{4\pi} \right)
\frac{(5+2b_0r^2)}{(1+2b_0 r^2)^3} \,.
 \nonumber
\end{eqnarray}

The anisotropy factor takes the following form
\begin{equation}
\Delta=\left(\frac{b_0^2r^2}{4\pi}\right)\frac{(5+2b_0
r^2)}{(1+2b_0 r^2)^3}\,,
\end{equation}
which is always positive, implying that $p_{\perp}>p_r$, and
$\Delta|_{r=0}=0$ at the center, i.e., $p_{\perp}(0)=p_r(0)$, as
expected.

\subsection{Slowly rotating gravastars}

When the equilibrium configuration described in Sec.
\ref{sec:graveqs} is set into slow rotation, the geometry of
spacetime around it and its interior distribution of stress-energy
are changed. With an appropriate change of coordinates, the
perturbed geometry is described by \cite{Ha67,HaTh68}
\begin{widetext}
\vspace{-0.5cm}

\begin{eqnarray}
ds^{2} &=&-e^{\nu _{rot}(r)}\left\{ 1+2\left[
h_{0}+h_{2}P_{2}\left( \cos \theta \right) \right] \right\}
dt^{2}+\frac{1+2\left[ m_{0}+m_{2}P_{2} \left( \cos \theta \right)
\right] /\left( r-2M\right) }{1-2M/r}dr^{2}+
\nonumber \\
&&r^{2}\left[ 1+2\left( v_{2}-h_{2}\right) P_{2}\left( \cos \theta
\right) \right] \left[ d\theta ^{2}+\sin ^{2}\theta \left(
d\varphi -\omega dt\right) ^{2}\right] +O\left( \Omega _S^{3}\right)
.
\end{eqnarray}

Here, $P_{2}\left( \cos \theta \right) =\left( 3\cos ^{2}\theta
-1\right) /2$ is the Legendre polynomial of order two; the angular
velocity, $\omega $, of the local inertial frame, is a
function of the radial coordinate $r$, and is proportional to the
star's angular velocity $\Omega _S$; and $h_{0}$, $h_{2}$, $m_{0}$,
$m_{2}$, $v_{2}$ are functions of $r$ that are proportional to
$\Omega _S^{2}$. Outside the star the metric can be written as
\cite{HaTh68}
\begin{eqnarray}\label{metric}
ds^{2} &=&-\left( 1-\frac{2M}{r}+2\frac{J^{2}}{r^{4}}\right)
\left\{ 1+2 \left[ \frac{J^{2}}{Mr^{3}}\left( 1+\frac{M}{r}\right)
+\frac{5}{8}\frac{ Q-J^{2}/M}{M^{3}}Q_{2}^{2}\left( \chi \right)
\right] P_{2}\left(
\cos \theta \right) \right\} dt^{2}   \nonumber\\
&&+\left( 1-\frac{2M}{r}+2\frac{J^{2}}{r^{4}}\right) ^{-1}\left\{
1-2\left[ \frac{J^{2}}{Mr^{3}}\left( 1-\frac{5M}{r}\right)
+\frac{5}{8}\frac{Q-J^{2}/M }{M^{3}}Q_{2}^{2}\left( \chi \right)
\right] P_{2}\left( \cos \theta
\right) \right\} dr^{2}  \nonumber \\
&&+r^{2}\left\{ 1+2\left[ -\frac{J^{2}}{Mr^{3}}\left(
1+\frac{2M}{r}\right) + \frac{5}{8}\frac{Q-J^{2}/M}{M^{3}}\left(
\frac{2M}{\sqrt{r\left( r-2M\right) }}Q_{2}^{1}\left( \chi\right)
-Q_{2}^{2}\left( \chi \right)
\right) \right]P_{2}\left( \cos \theta \right) \right\} \nonumber\\
&& \times \left\{ d\theta ^{2}+\sin ^{2}\theta \left[ d\varphi
-\frac{2J}{ r^{3}}dt\right] ^{2}\right\} .
\end{eqnarray}

\end{widetext}

In Eq.~(\ref{metric}) the variable $\chi =r/M-1$, and the
quantities $Q_{2}^{1}$ and $Q_{2}^{2}$ denote associated Legendre
polynomials of the second time, so that
\begin{equation}
Q_{2}^{1}(\chi )=\sqrt{\chi ^{2}-1}\left[\frac{\left( 3\chi ^{2}
-2\right)}{\chi ^{2}-1}-\frac{3}{2%
}\chi \ln \frac{\chi +1}{\chi -1}\right],
\end{equation}
and
\begin{equation}
Q_{2}^{2}(\chi )=\frac{  5\chi
-3\chi ^{2} }{\left(
\chi ^{2}-1\right)}+\frac{3}{2}\left( \chi ^{2}-1\right)
\ln \frac{\chi +1}{\chi -1},
\end{equation}
respectively. The line element outside the star is determined by
three constants: the total mass of the rotating star $M$, the
star's total angular momentum $J$ and the star's mass quadrupole
moment $Q$. The quadrupole moment $ Q$ can be expressed in terms
of the eccentricity $e=\sqrt{\left( r_{e}/r_{p}\right) ^{2}-1}$ of
the star as $e=\sqrt{3Q/Mr^{\ast 2}}+O\left( 1/r^{\ast 2}\right)
$, where $r^{\ast }$ is a large distance from the origin, and
$r_{e}$ and $r_{p}$ are the equatorial and polar radius of the
star, respectively \cite{HaTh68}. The metric given by
Eq.~(\ref{metric}) is valid in the case of slow rotation, that is,
the angular velocity of the star $\Omega _S$ must be small enough
so that the fractional changes in pressure, energy density, and
gravitational field, due to rotation, are all much less than
unity. The condition of slow rotation can be formulated as $\Omega
_S^2\ll\left(c/R\right)^2\left(GM/Rc^2\right)$, where $R$ is the
radius of the static stellar configuration \cite{Ha67}. The
critical angular velocity at which mass shedding occurs is given
by $\Omega _K^2=GM/R^3$ \cite{HaTh68}, and therefore the condition
of slow rotation implies $\Omega _S\ll\Omega _K$. All models
considered in the present paper satisfy this condition.

The metric given by Eq. (\ref{metric}) is used to determine the
electromagnetic signatures of accretion disks around slowly
rotating gravastars, which is analyzed in detail below.

\subsection{The Kerr black hole}

For self-completeness and self-consistency, we present the Kerr
metric, as it will be compared to metric (\ref{metric}) in the
electromagnetic signature analysis of accretion disks. The Kerr
metric, describing a rotating black hole, is given in the
Boyer-Lyndquist coordinate system by
\begin{eqnarray}
ds^{2} &=&-\left( 1-\frac{2Mr}{\Sigma _K}\right)
dt^{2}+2\frac{2Mr}{\Sigma _K} a\sin ^{2}\theta dtd\phi +\frac{\Sigma _K
}{\Delta _K}dr^{2}
    \nonumber \\
&& \hspace{-0.5cm}+\Sigma _Kd\theta ^{2}+\left(
r^{2}+a^{2}+\frac{2Mr}{\Sigma _K} a^{2}\sin ^{2}\theta \right) \sin
^{2}\theta d\phi ^{2},
\end{eqnarray}
where $\Sigma _K=r^2+a^2\cos ^2\theta$ and $\Delta _K=r^2+a^2-2mr$, respectively.
In the equatorial plane, the metric components reduce to
\begin{eqnarray*}
g_{tt} &=&-\left( 1-\frac{2Mr}{\Sigma _K}\right) =-\left(
1-\frac{2M}{r}
\right) \;, \\
g_{t\phi } &=&\frac{2Mr}{\Sigma _K}a\sin ^{2}\theta =2\frac{Ma}{r}\;, \\
g_{rr} &=&\frac{\Sigma _K}{\Delta _K}=\frac{r^{2}}{\Delta _K}\;, \\
g_{\phi \phi } &=&\left( r^{2}+a^{2}+\frac{2Mr}{\Sigma _K}a^{2}\sin
^{2}\theta \right) \sin ^{2}\theta
    \nonumber   \\
&=&r^{2}+a^{2}\left( 1+\frac{2M}{r}\right) \,,
\end{eqnarray*}%
respectively. For the Kerr metric $J=-Ma$ and $Q=J^{2}/M$,
respectively. The latter relationship, i.e., $Q=J^{2}/M$, between
the quadrupole momentum and the angular momentum shows the very
special nature of the Kerr solution.

\section{Electromagnetic radiation properties of thin accretion disks in
stationary axisymmetric spacetimes}\label{sec:III}

To set the stage, we present the general formalism of
electromagnetic radiation properties of thin accretion disks in
stationary axisymmetric spacetimes.

\subsection{Stationary and axially symmetric spacetimes}

In this work we analyze the physical properties and
characteristics of particles moving in circular orbits around
general relativistic compact spheres in a stationary and axially
symmetric geometry given by the following metric
\begin{equation}  \label{rotmetr1}
ds^2=g_{tt}\,dt^2+2g_{t\phi}\,dt d\phi+g_{rr}\,dr^2
+g_{\theta\theta}\,d\theta^2+g_{\phi\phi}\,d\phi^2\,.
\end{equation}
Note that the metric functions $g_{tt}$, $g_{t\phi}$, $g_{rr}$, $
g_{\theta\theta}$ and $g_{\phi\phi}$ only depend on the radial
coordinate $r$ in the equatorial approximation, i.e.,
$|\theta-\pi|\ll 1$, which is the case of interest here. In the
following we denote the square root of the determinant of the
metric tensor by $\sqrt{-g}$.

To compute the flux integral given by Eq.~(\ref{F}), we determine
the radial dependence of the angular velocity $\Omega $, of the
specific energy $ \widetilde{E}$ and of the specific angular
momentum $\widetilde{L}$ of particles moving in circular orbits
around compact spheres in a stationary and axially symmetric
geometry through the geodesic equations. The latter take the
following form
\begin{eqnarray}
\frac{dt}{d\tau}&=&\frac{\widetilde{E}
g_{\phi\phi}+\widetilde{L}g_{t\phi}}{
g_{t\phi}^2-g_{tt}g_{\phi\phi}}\,,  \label{geodeqs1} \\
\frac{d\phi}{d\tau}&=&-\frac{\widetilde{E}
g_{t\phi}+\widetilde{L}g_{tt}
}{g_{t\phi}^2-g_{tt}g_{\phi\phi}}\,,  \label{geodeqs2} \\
g_{rr}\left(\frac{dr}{d\tau}\right)^2&=&-1+\frac{\widetilde{E}^2
g_{\phi\phi}+2\widetilde{E}\widetilde{L}g_{t\phi}
+\widetilde{L}^2g_{tt}}{ g_{t\phi}^2-g_{tt}g_{\phi\phi}}\,.
\label{geodeqs3}
\end{eqnarray}
One may define an effective potential term defined as
\begin{equation}  \label{roteffpot}
V_{eff}(r)=-1+\frac{\widetilde{E}^2
g_{\phi\phi}+2\widetilde{E}\widetilde{L} g_{t\phi}
+\widetilde{L}^2g_{tt}}{g_{t\phi}^2-g_{tt}g_{\phi\phi}}\,.
\end{equation}

For stable circular orbits in the equatorial plane the following
conditions must hold: $V_{eff}(r)=0$ and $V_{eff,\;r}(r)=0$. These
conditions provide the specific energy, the specific angular
momentum and the angular velocity of particles moving in circular
orbits for the case of spinning general relativistic compact
spheres, given by
\begin{eqnarray}
\widetilde{E}&=&-\frac{g_{tt}+g_{t\phi}\Omega}{\sqrt{-g_{tt}
-2g_{t\phi}\Omega-g_{\phi\phi}\Omega^2}}\,,  \label{rotE} \\
\widetilde{L}&=&\frac{g_{t\phi}+g_{\phi\phi}\Omega}{\sqrt{-g_{tt}
-2g_{t\phi}\Omega-g_{\phi\phi}\Omega^2}}\,,  \label{rotL} \\
\Omega&=&\frac{d\phi}{dt}=\frac{-g_{t\phi,r}+\sqrt{(g_{t\phi,r})^2
-g_{tt,r}g_{\phi\phi,r}}}{g_{\phi\phi,r}}\,.  \label{rotOmega}
\end{eqnarray}
The marginally stable orbit around the central object can be
determined from the condition $V_{eff,\;rr}(r)=0$, which provides
the following important relationship
\begin{eqnarray}
\widetilde{E}^{2}g_{\phi\phi,rr}+2\widetilde{E}\widetilde{L}
g_{t\phi,rr}+ \widetilde{L}^{2}g_{tt,rr} -(g_{t\phi}^{2}
-g_{tt}g_{\phi\phi})_{,rr} =0.\nonumber\\  \label{mso-r}
\end{eqnarray}
By inserting Eqs. (\ref{rotE})-(\ref{rotOmega}) into Eq.
(\ref{mso-r}) and solving this equation for $r$, we obtain the
marginally stable orbit for the explicitly given metric
coefficients $g_{tt}$, $g_{t\phi}$ and $g_{\phi\phi}$. For a Kerr
black hole the geodesic equation (\ref{geodeqs3}) for $r$ becomes
\begin{equation}
\frac{r^{2}}{\Delta _K}\left(\frac{dr}{d\tau}\right)^{2}=V_{eff}(r)
\end{equation}
with the effective potential given by
\begin{eqnarray}
V_{eff}(r)= -1+\Big\{\widetilde{E}^{2}\left[r^{2}(r^{2}+a^{2})
+2ma^{2}r \right]
   \nonumber   \\
+4\widetilde{E}\widetilde{L}mar-\widetilde{L}^{2}
\left(r^{2}-2mr\right)\Big\}\big/\left[
r^{2}\left(r^2-2mr+a^2\right)\right]. \nonumber\\
\end{eqnarray}

Note that these relationships may be rewritten in the following
manner
\begin{equation}
r^4\left(\frac{dr}{d\tau}\right)^{2}=V(r)  \label{KerrPot}
\end{equation}
with $V(r)$ given by
\begin{equation}
V(r)=r^{2}\Delta _KV_{eff}(r)=r^{2}(r^2-2mr+a^2)
V_{eff}(r)\;.  \label{Veff}
\end{equation}
where the relationship $\Delta _K=g_{t\phi}^{2}-g_{tt}g_{\phi\phi}=r^2-2mr+a^2$
along the equatorial plane has been used.

\subsection{Properties of thin accretion disks}

For the thin accretion disk, it is assumed that its vertical size
is negligible, as compared to its horizontal extension, i.e, the
disk height $H$, defined by the maximum half thickness of the disk
in the vertical direction, is always much smaller than the
characteristic radius $r$ of the disk, defined along the
horizontal direction, $H \ll r$. The thin disk is in
hydrodynamical equilibrium, and the pressure gradient and a
vertical entropy gradient in the accreting matter are negligible.
The efficient cooling via the radiation over the disk surface
prevents the disk from cumulating the heat generated by stresses
and dynamical friction. In turn, this equilibrium causes the disk
to stabilize its thin vertical size. The thin disk has an inner
edge at the marginally stable orbit of the compact object
potential, and the accreting plasma has a Keplerian motion in
higher orbits.

In steady state accretion disk models, the mass accretion rate
$\dot{M}_{0}$ is assumed to be a constant that does not change
with time. The physical quantities describing the orbiting plasma
are averaged over a characteristic time scale, e.g. $\Delta t$,
over the azimuthal angle $\Delta \phi =2\pi $ for a total period
of the orbits, and over the height $H$ \cite{ShSu73,
NoTh73,PaTh74}.

The particles moving in Keplerian orbits around the compact object
with a rotational velocity $\Omega =d\phi /dt$ have a specific
energy $\widetilde{E} $ and a specific angular momentum
$\widetilde{L\text{,}}$ which, in the steady state thin disk
model, depend only on the radii of the orbits. These particles,
orbiting with the four-velocity $u^{\mu }$, form a disk of an
averaged surface density $\Sigma $, the vertically integrated
average of the rest mass density $\rho _{0}$ of the plasma. The
accreting matter in the disk is modeled by an anisotropic fluid
source, where the density $\rho _{0}$ , the energy flow vector
$q^{\mu }$ and the stress tensor $t^{\mu \nu }$ are measured in
the averaged rest-frame (the specific heat was neglected). Then,
the disk structure can be characterized by the surface density of
the disk \cite{NoTh73,PaTh74},
\begin{equation}
\Sigma(r) = \int^H_{-H}\langle\rho_0\rangle dz,
\end{equation}
with averaged rest mass density $\langle\rho_0\rangle$ over
$\Delta t$ and $ 2\pi$ and the torque
\begin{equation}
W_{\phi}{}^{r} =\int^H_{-H}\langle t_{\phi}{}^{r}\rangle dz,
\end{equation}
with the averaged component $\langle t^r_{\phi} \rangle$ over
$\Delta t$ and $2\pi$. The time and orbital average of the energy
flux vector gives the radiation flux ${\mathcal{F}}(r)$ over the
disk surface as ${\mathcal{F}} (r)=\langle q^z \rangle$.

The stress-energy tensor is decomposed according to
\begin{equation}
T^{\mu \nu }=\rho_{0}u^{\mu }u^{\nu }+2u^{(\mu }q^{\nu )}+t^{\mu \nu },
\end{equation}
where $u_{\mu }q^{\mu }=0$, $u_{\mu }t^{\mu \nu }=0$. The
four-vectors of the energy and angular momentum flux are defined
by $-E^{\mu }\equiv T_{\nu }^{\mu }{}(\partial /\partial t)^{\nu
}$ and $J^{\mu }\equiv T_{\nu }^{\mu }{}(\partial /\partial \phi
)^{\nu }$, respectively. The structure equations of the thin disk
can be derived by integrating the conservation laws of the rest
mass, of the energy, and of the angular momentum of the plasma,
respectively \cite{NoTh73,PaTh74}. From the equation of the rest
mass conservation, $\nabla _{\mu }(\rho _{0}u^{\mu })=0$, it
follows that the time averaged rate of the accretion of the rest
mass is independent of the disk radius,
\begin{equation}
\dot{M_{0}}\equiv -2\pi \sqrt{-g}\Sigma u^{r}=\mathrm{constant}.
\label{conslawofM}
\end{equation}

The conservation law $\nabla _{\mu }E^{\mu }=0$ of the energy has
the integral form
\begin{equation}
\lbrack \dot{M}_{0}\widetilde{E}-2\pi \sqrt{-g}\Omega W_{\phi
}{}^{r}]_{,r}=4\pi \sqrt{-g}F \widetilde{E}\;\;,  \label{conslawofE}
\end{equation}%
which states that the energy transported by the rest mass flow,
$\dot{M}_{0} \widetilde{E}$, and the energy transported by the
dynamical stresses in the disk, $2\pi \sqrt{-g}\Omega W_{\phi }{}^{r}$, is
in balance with the energy radiated away from the surface of the
disk, $4\pi \sqrt{-g}F\widetilde{E}$. The law of the angular momentum
conservation, $\nabla _{\mu }J^{\mu }=0$, also states the balance
of these three forms of the angular momentum transport,
\begin{equation}
\lbrack \dot{M}_{0}\widetilde{L}-2\pi rW_{\phi }{}^{r}]_{,r}=4\pi
\sqrt{-g}F \widetilde{L}\;\;.  \label{conslawofL}
\end{equation}

By eliminating $W_{\phi }{}^{r}$ from Eqs. (\ref{conslawofE}) and
(\ref {conslawofL}), and applying the universal energy-angular
momentum relation $ dE=\Omega dJ$ for circular geodesic orbits in
the form $\widetilde{E} _{,r}=\Omega \widetilde{L}_{,r}$, the flux
$F$ of the radiant energy over the disk can be expressed in terms
of the specific energy, angular momentum and of the angular
velocity of the central compact object \cite{NoTh73,PaTh74},
\begin{equation}
F(r)=-\frac{\dot{M}_{0}}{4\pi \sqrt{-g}}\frac{\Omega
_{,r}}{(\widetilde{E}-\Omega
\widetilde{L})^{2}}\int_{r_{ms}}^{r}(\widetilde{E}-\Omega
\widetilde{L}) \widetilde{L}_{,r}dr\;\;.  \label{F}
\end{equation}

Another important characteristics of the mass accretion process is
the efficiency with which the central object converts rest mass
into outgoing radiation. This quantity is defined as the ratio of
the rate of the radiation of energy of photons escaping from the
disk surface to infinity, and the rate at which mass-energy is
transported to the central compact general relativistic object,
both measured at infinity \cite{NoTh73,PaTh74}. If all the emitted
photons can escape to infinity, the efficiency is given in terms
of the specific energy measured at the marginally stable orbit $
r_{ms}$,
\begin{equation}
\epsilon =1-\widetilde{E}_{ms}.  \label{epsilon}
\end{equation}
For Schwarzschild black holes the efficiency $\epsilon $ is about
$6\%$, whether the photon capture by the black hole is considered,
or not. Ignoring the capture of radiation by the hole, $\epsilon $
is found to be $42\%$ for rapidly rotating black holes, whereas
the efficiency is $40\%$ with photon capture in the Kerr potential
\cite{Th74}.

The accreting matter in the steady-state thin disk model is
supposed  to be in thermodynamical equilibrium. Therefore the
radiation emitted by the disk surface can be considered as a
perfect black body radiation, where the energy flux is given by
$F(r)=\sigma _{SB}T^{4}(r)$ ($\sigma _{SB}$ is the Stefan-Boltzmann
constant), and the observed luminosity $L\left( \nu \right)$ has a
redshifted black body spectrum \citep{To02}:
\begin{equation}
L\left( \nu \right) =4\pi d^{2}I\left( \nu \right) =\frac{8}{\pi  }\cos
\gamma \int_{r_{i}}^{r_{f}}\int_0^{2\pi}\frac{\nu^{3}_e r d\phi dr }{\exp
\left(\nu_e/T\right) -1}.
\end{equation}

Here $d$ is the distance to the source, $I(\nu )$ is the Planck
distribution function, $\gamma $ is the disk inclination angle,
and $r_{i}$ and $r_{f}$ indicate the position of the inner and
outer edge of the disk, respectively. We take $r_{i}=r_{ms}$ and
$r_{f}\rightarrow \infty $, since we expect the flux over the disk
surface vanishes at $r\rightarrow \infty $ for any kind of general
relativistic compact object geometry. The emitted frequency is
given by $\nu_e=\nu(1+z)$, where the redshift factor can be
written as
\begin{equation}
1+z=\frac{1+\Omega r \sin \phi \sin \gamma }{\sqrt{ -g_{tt}-2
\Omega g_{t\phi} - \Omega^2 g_{\phi\phi}}}
\end{equation}
where we have neglected the light bending \cite{Lu79,BMT01}.

\section{Electromagnetic and thermodynamic signatures of accretion
disks around slowly rotating gravastars and black holes}\label{sec:IV}

\subsection{Electromagnetic and thermodynamic properties of the disks}

\begin{figure*}
\centering
  \includegraphics[width=2.65in]{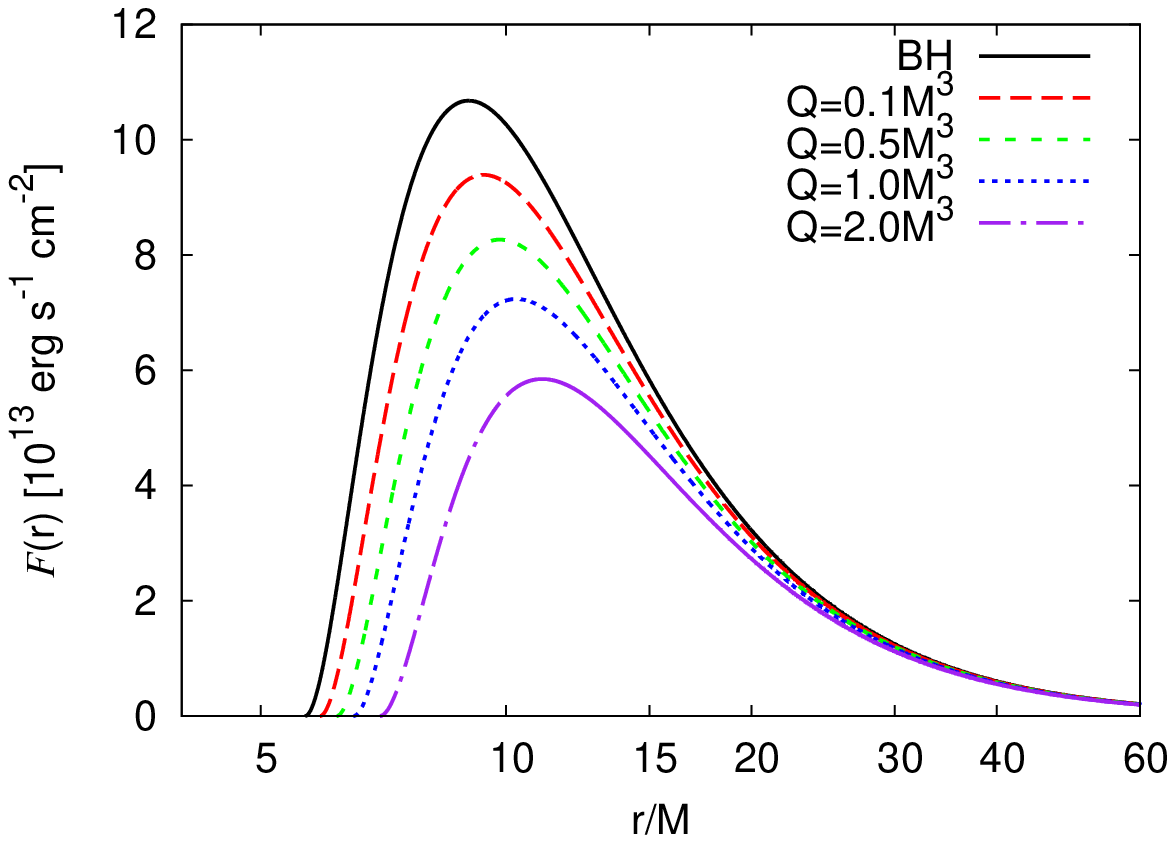}
\hspace{0.2in}
  \includegraphics[width=2.65in]{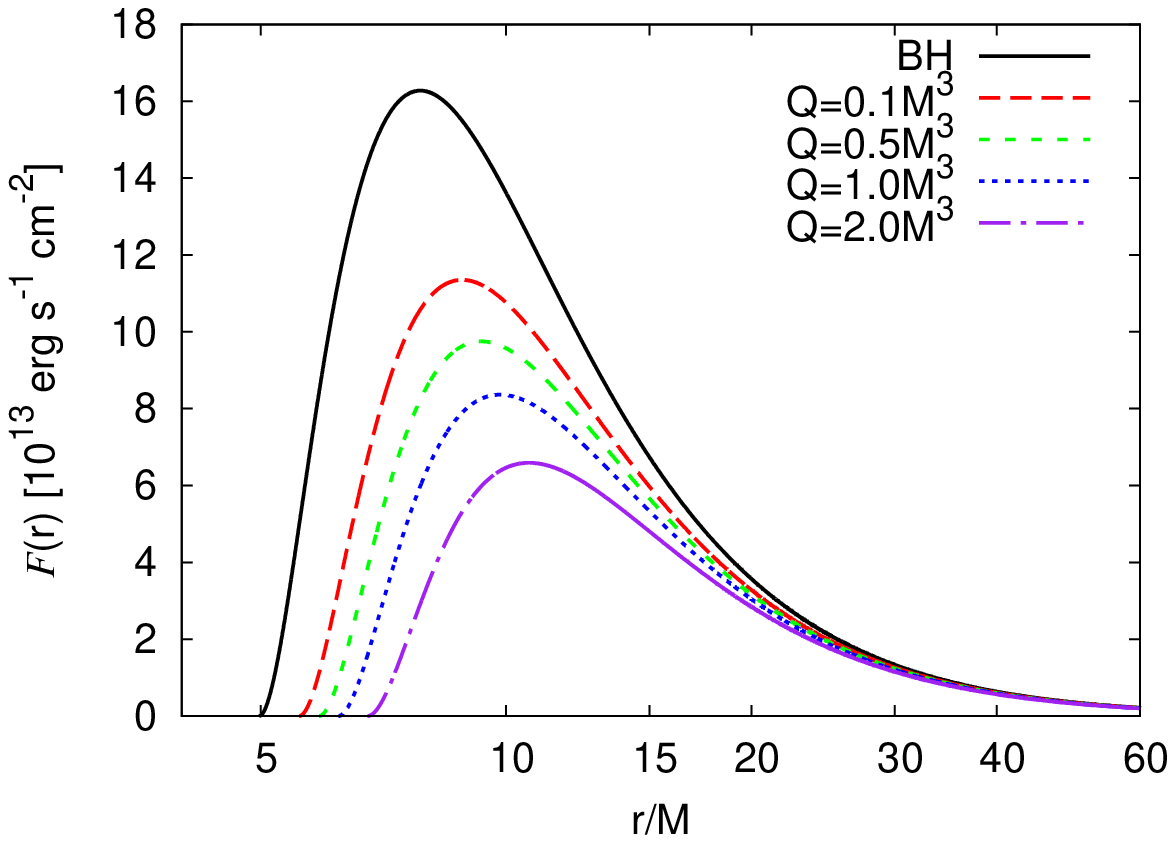}
\hspace{0.2in}
  \includegraphics[width=2.65in]{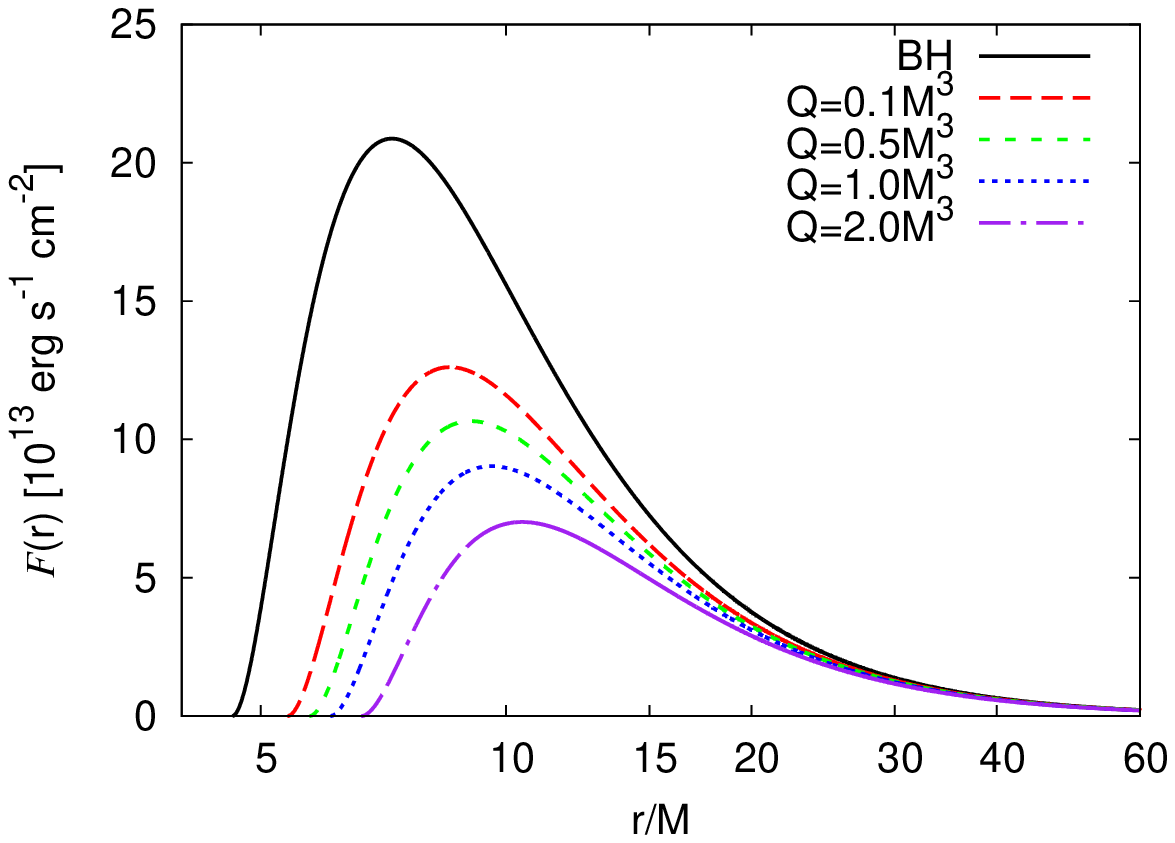}
\hspace{0.2in}
   \includegraphics[width=2.65in]{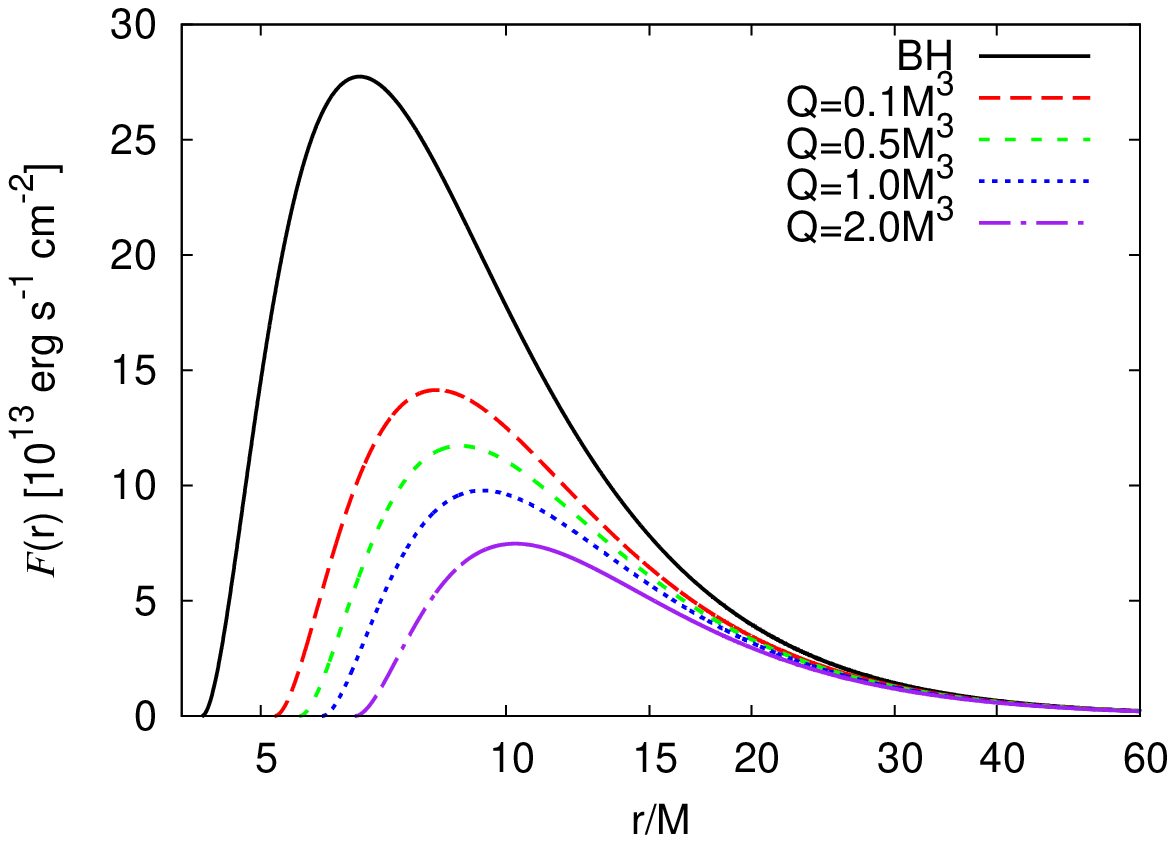}
\caption{The energy flux emerging from the accretion disk of
slowly rotating gravastars and black holes for the spin parameter
$a_*=0.1$ (upper left hand plot), $a_*=0.3$ (upper right hand
plot), $a_*= 0.4$ (lower left hand plot), and $a_*=0.5$ for (lower
right hand plot). All the plots are given for the total mass $M =
10^6 M_{\odot}$, the quadrupole moments $Q = 0.1, 0.5, 1.0, 2.0$
times $M^3$, and the mass accretion rate $2.5\times10^{-5}
M_{\odot}$/yr.}
 \label{Fig:flux}
\end{figure*}

In this section we compare the radiation properties of thin
accretion disks around gravastars and black holes in the slowly
rotating case when the spin parameter $a_*=J/M^2$ has the maximal
value of 0.5. In Fig.~\ref{Fig:flux} we present the time averaged
energy flux $F(r)$ radiated by the disk for both types of central
objects with the total mass $M$ of $10^6M_{\odot}$ and increasing
spin parameter from 0.1 to 0.5. The quadrupole moment $Q$ of the
gravastar models runs between $0.1M^3$ and $2M^3$. Here the mass
accretion rate $\dot{M}_{0}$ is set to $2.5\times10^{-5}
M_{\odot}$/yr, which is in the typical range for super massive
cental objects.

If we compare the flux emerging from the surface of the thin
accretion disk around black holes and gravastars, we find that its
maximal value is systematically lower for gravastars,
independently of the values of the spin parameter or the
quadrupole momentum. For very slow rotation, ($a_*=0.1$), and a
relative small value of the quadrupole moment ($Q=0.1M^3$), the
radial distribution of the disk radiation is close to each other
for the two types of compact central objects. The maximal flux for
gravastars is roughly 90\% of the black hole's flux, and the
maximum of the inner edge of the accretion disk is located at
somewhat higher radii for gravastars. With increasing rotational
frequency of the central object, the flux values also increase, but
the increment is higher for black holes than for gravastars. For
$a_*=0.5$ the flux maximum for black holes is almost twice the
maximal flux value for gravastars. The more rapid rotation does
not cause strong effect on the location of the inner disk edge for
gravastars, as we find a slight decrease in the value of $r_{ms}$
as the rotation of central object increases (see in Tab
\ref{Tab:eff_gs}). For black holes, the effect of the rotation is
also stronger here. Although the error of the approximation
applied for the slow rotation is rapidly increasing in the regime
around $a_*\sim 0.5$, this picture is still definitely adequate
for lower values of $a_*$.

\begin{figure*}
\centering
  \includegraphics[width=2.65in]{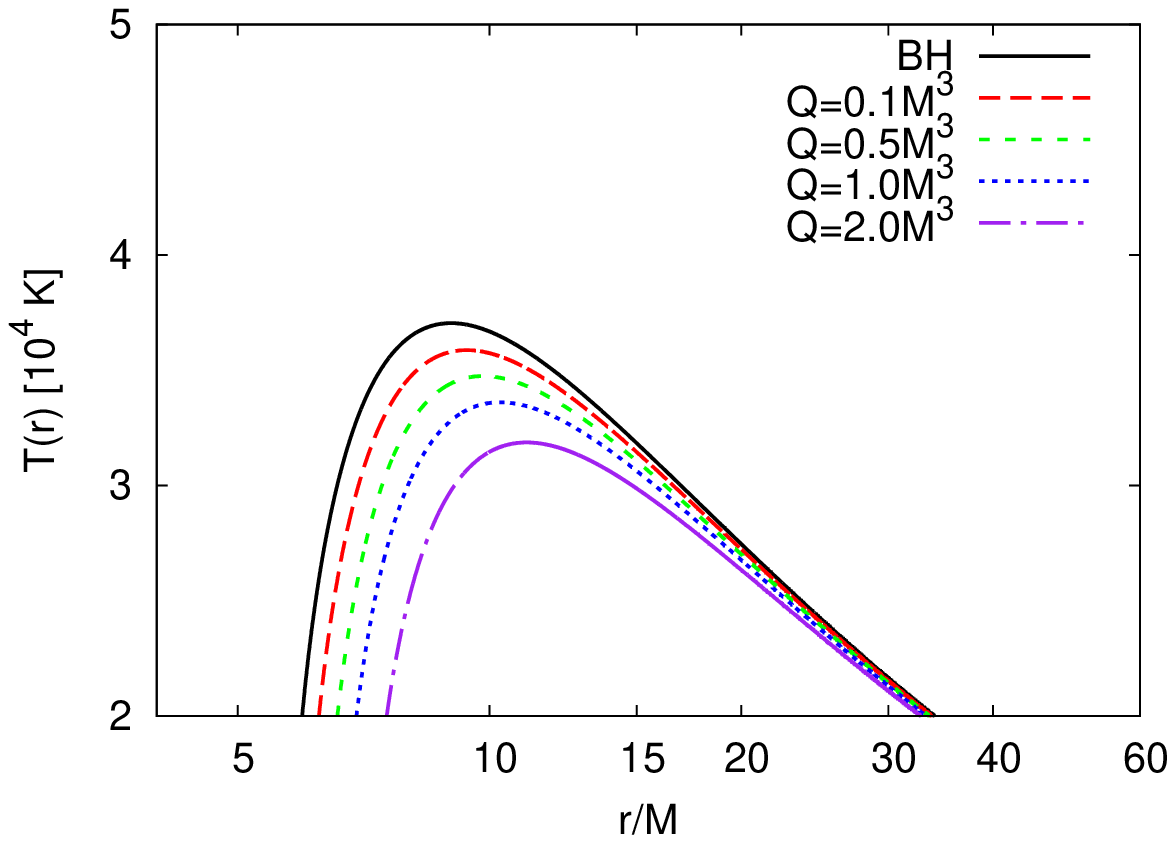}
\hspace{0.1in}
  \includegraphics[width=2.65in]{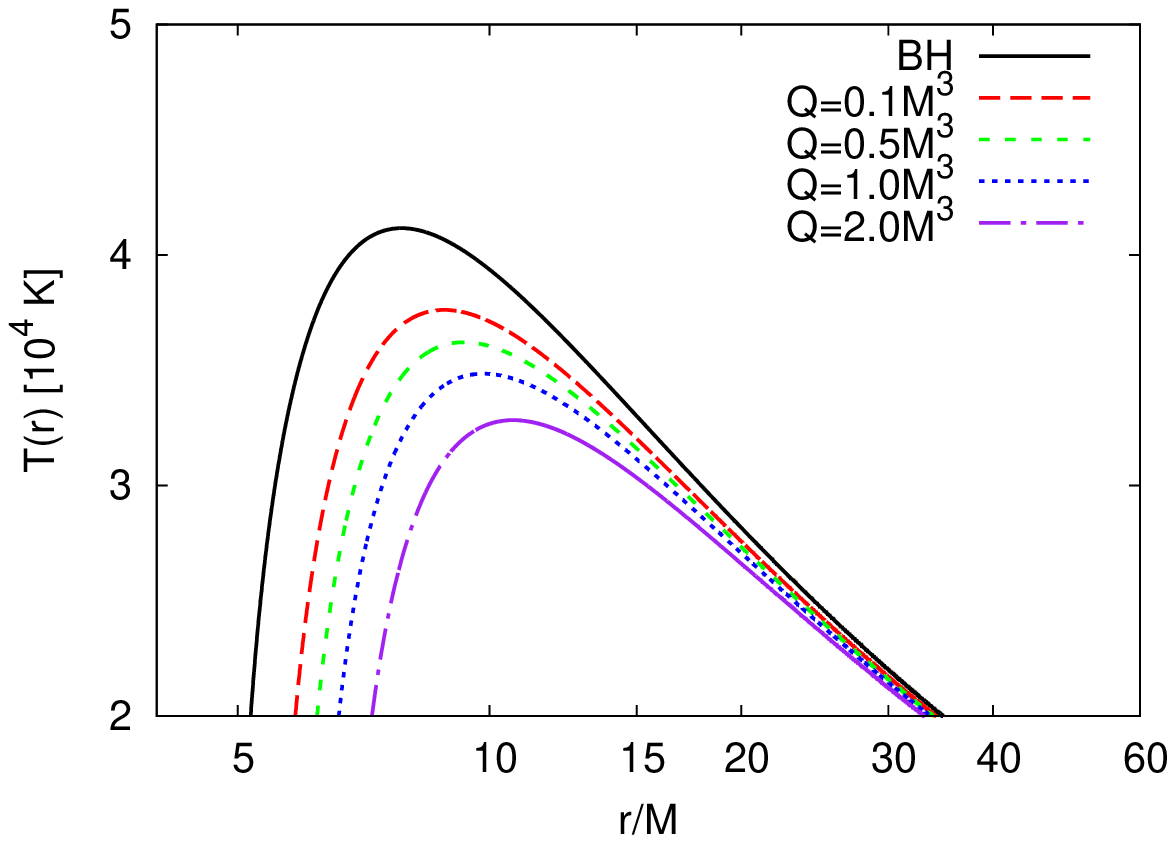}
\hspace{0.1in}
  \includegraphics[width=2.65in]{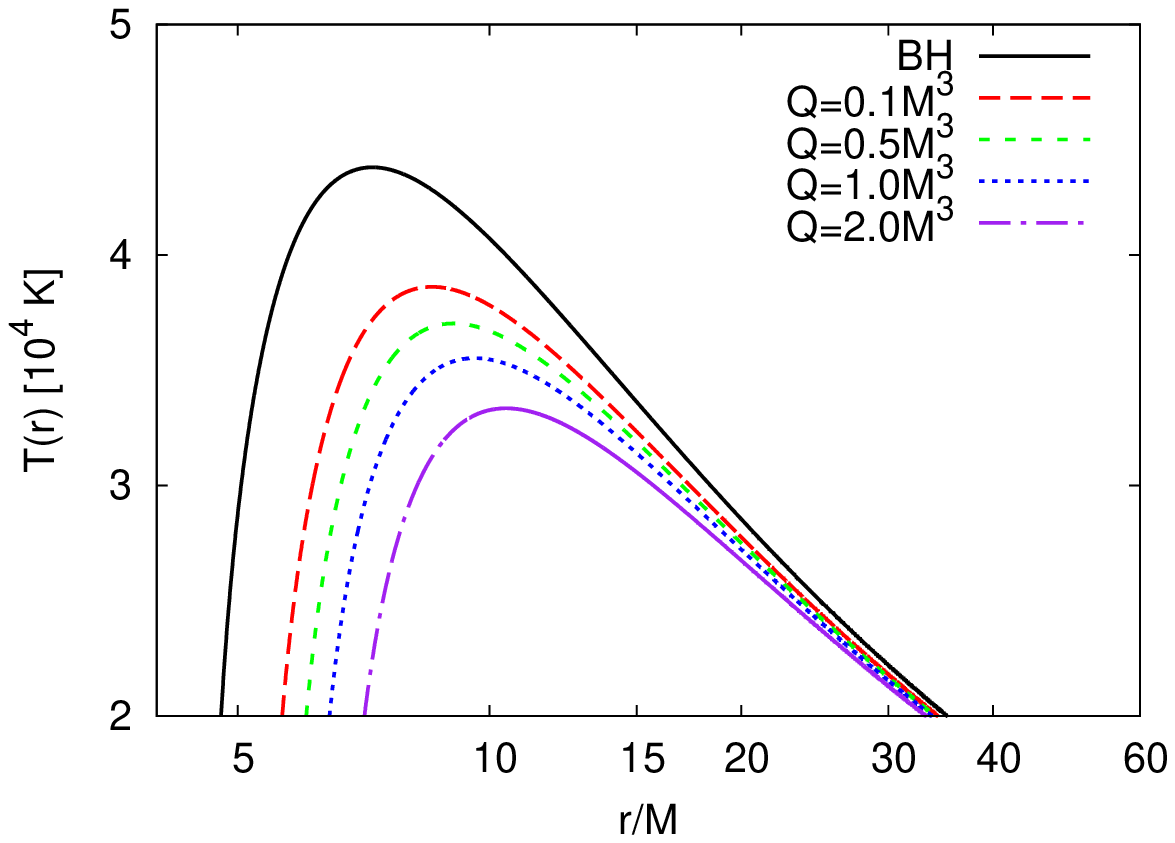}
\hspace{0.1in}
   \includegraphics[width=2.65in]{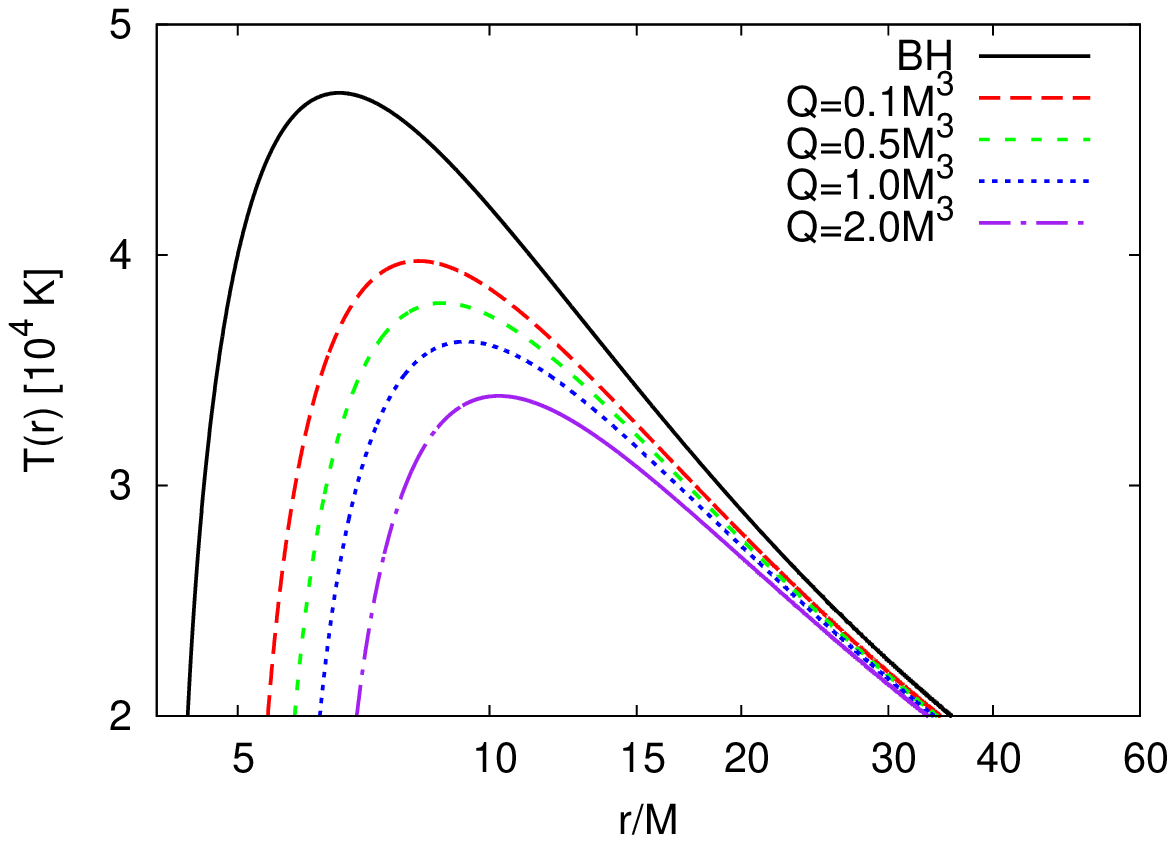}
\caption{The disk temperature of slowly rotating gravastars and
black holes for the spin parameter $a_*=0.1$ (upper left hand
plot), $a_*=0.3$ (upper right hand plot), $a_*= 0.4$ (lower left
hand plot), and $a_*=0.5$ for (lower right hand plot). All the
plots are given for the total mass $M = 10^6 M_{\odot}$, the
quadrupole moments $Q = 0.1, 0.5, 1.0, 2.0$ times $M^3$, and the
mass accretion rate $2.5\times10^{-5} M_{\odot}$/yr.}
 \label{Fig:temp}
\end{figure*}

The variation of the quadrupole moment causes considerable
changes in both the maximal value of disk radiation and the
location of the inner edge of the disk. As we increase $Q$, the
maximal flux decreases and $r_{ms}$ increases. These effects are
presented in Fig.~\ref{Fig:temp}, showing the disk temperatures,
although the differences are somewhat less striking.

\begin{figure*}
\centering
  \includegraphics[width=2.65in]{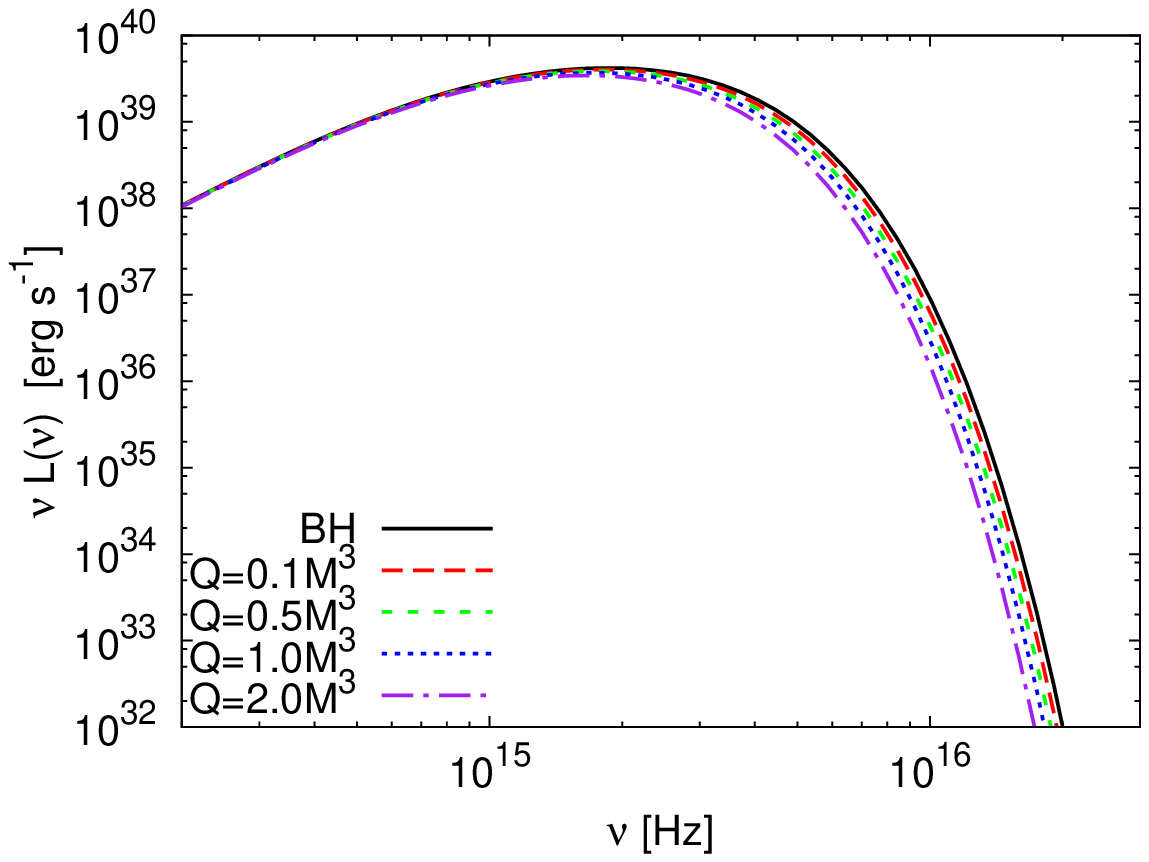}
\hspace{0.1in}
  \includegraphics[width=2.65in]{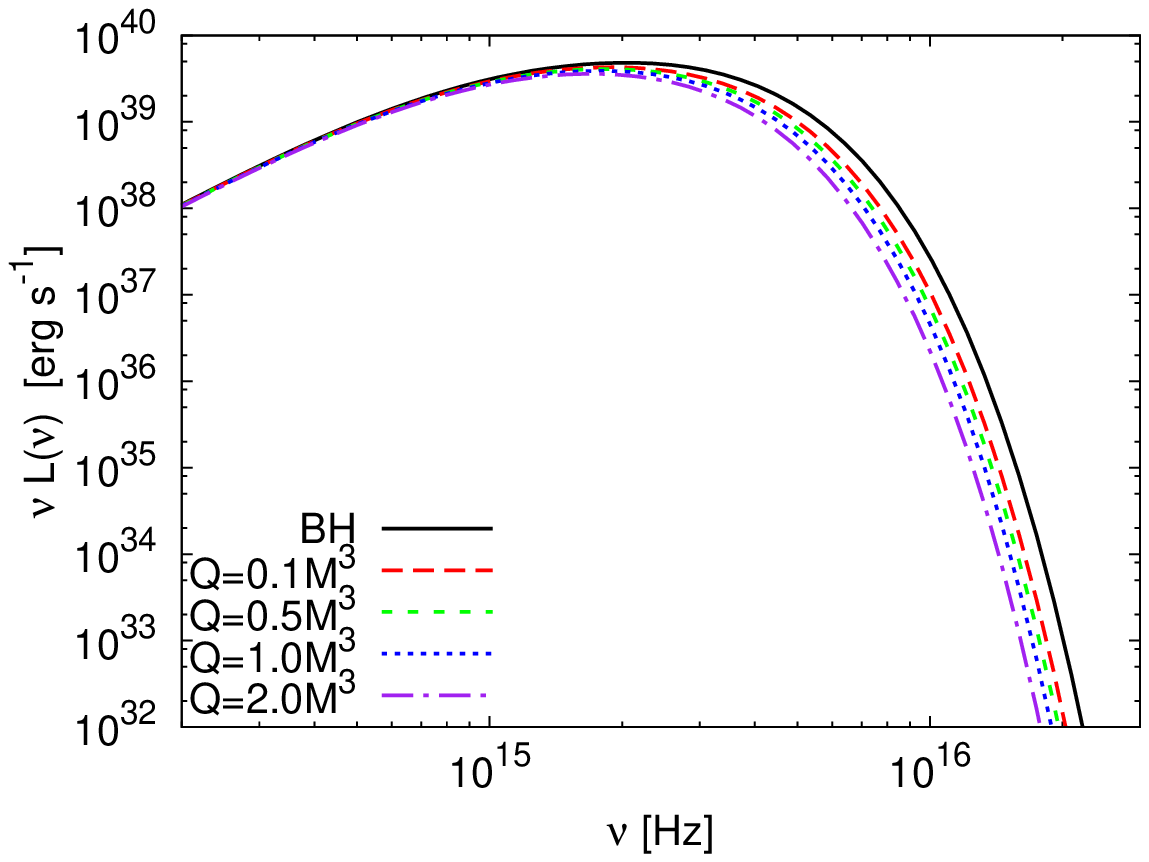}
\hspace{0.1in}
  \includegraphics[width=2.65in]{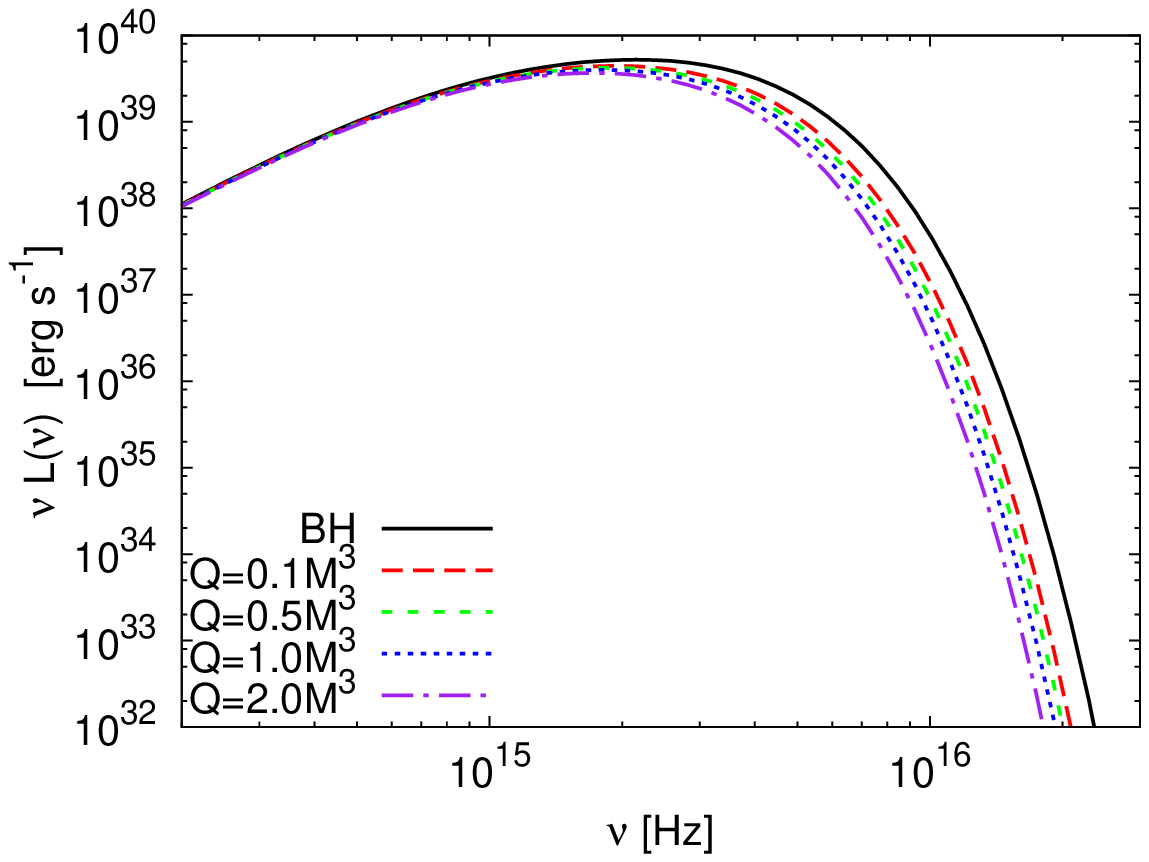}
\hspace{0.1in}
   \includegraphics[width=2.65in]{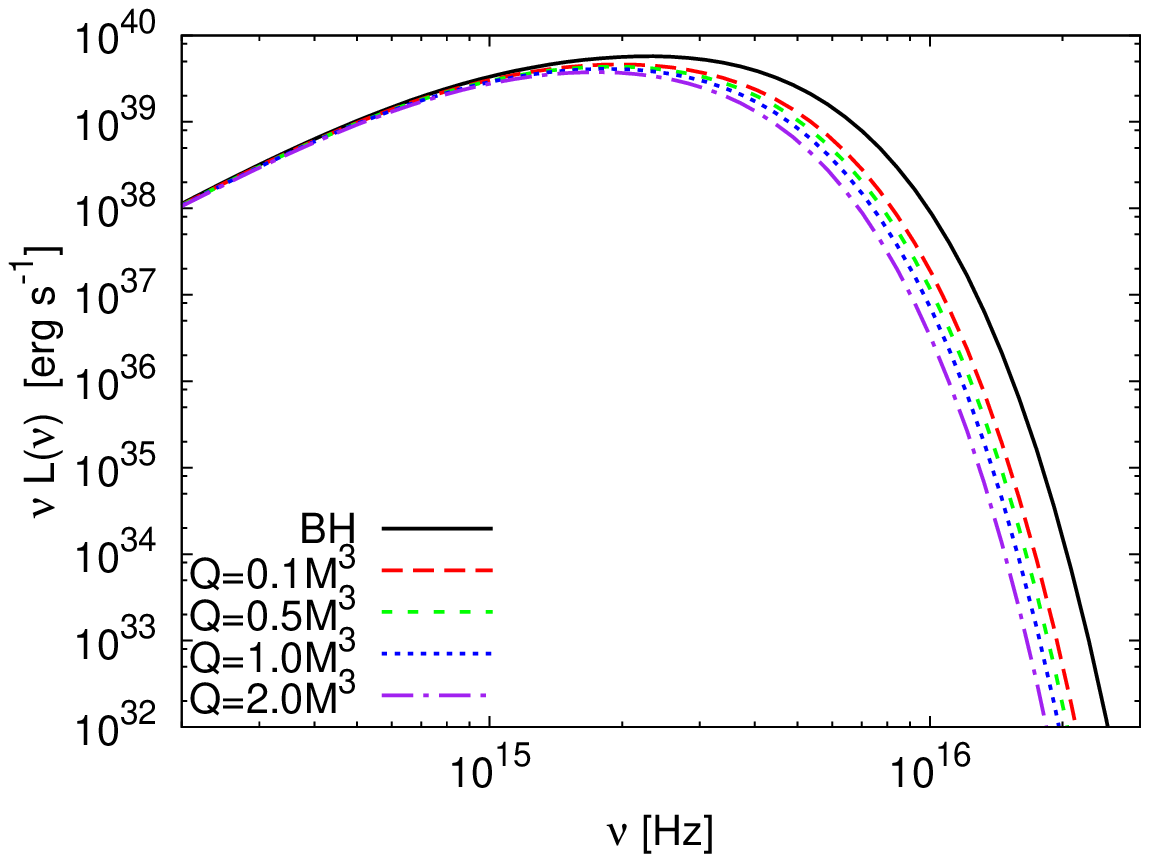}
\caption{The disk spectra of slowly rotating gravastars and black
holes for the spin parameter $a_*=0.1$ (upper left hand plot),
$a_*=0.3$ (upper right hand plot), $a_*= 0.4$ (lower left hand
plot), and $a_*=0.5$ for (lower right hand plot). All the plots
are given for the total mass $M = 10^6 M_{\odot}$, the quadrupole
moments $Q = 0.1, 0.5, 1.0, 2.0$ times $M^3$, and the mass
accretion rate $2.5\times10^{-5} M_{\odot}$/yr.}
 \label{Fig:lumin}
\end{figure*}

The disk spectra, presented in Fig.~\ref{Fig:lumin}, have similar
features in the dependence of the disk radiation on the rotation
parameter and the quadrupole momentum. The amplitudes and the
cut-off frequencies of the spectra for gravastars are always lower
than those for black holes. For higher rotational velocity, the
amplitudes are somewhat higher but do not exhibit much change. The
cut-off frequency for black holes increases moderately, whereas it
has only a negligible increment for gravastars. This makes the
differences in the spectral properties more acute for higher
values of the spin parameter ($a_*\gtrsim0.3$). The increase in
the quadrupole moment somewhat lowers the amplitude of the spectra
but causes a stronger decrease in the cut-off frequencies.

\subsection{Conversion efficiency of the
accreting mass}

We also present the conversion efficiency $\epsilon$ of the
accreting mass into radiation, measured at infinity, which is
given by Eq. (\ref{epsilon}), for the case where the photon
capture by the slowly rotating central object is ignored. The
value of $\epsilon$ measures the efficiency of energy generating
mechanism by mass accretion. The amount of energy released by
matter leaving the marginally stable orbit, and falling down the
black hole, is the binding energy $\widetilde{E}_{ms}$ of the
black hole potential. In Tabs.~\ref{Tab:eff_Kerr} and
\ref{Tab:eff_gs}, the marginally stable orbits $r_{ms}$ and
$\epsilon$ are given for black holes and gravastars with the
parameters $a_*$ and $Q$ in the range used for the plots
presenting the radiation properties of the accretion disks.

\begin{table}[h]
\begin{center}
\begin{tabular}{|c|c|c|}
\hline
  $a_*$ &  $r_{ms}$ [$M$] & $\epsilon$ [$10^{-2}$]\\
\hline
0.1 & 5.67 & 6.06 \\
\hline
0.2 & 5.33 & 6.46 \\
\hline
0.3 & 4.98 & 6.94 \\
\hline
0.4 & 4.62 & 7.51 \\
\hline
0.5 & 4.24 & 8.21 \\
\hline
\end{tabular}
\end{center}
\caption{The marginally stable orbit and the efficiency for slowly
rotating Kerr black holes with different spin parameters.}
\label{Tab:eff_Kerr}
\end{table}
\begin{table}[tbp]
\begin{center}
\begin{tabular}{|c|c|c|c|}
\hline
  $a_*$ & $Q$ $[M^3$] & $r_{ms}$ [$M$] & $\epsilon$ [$10^{-2}$]\\
\hline
0.1 & 0.1 & 5.91 & 5.82 \\
0.1 & 0.5 & 6.20 & 5.60 \\
0.1 & 1.0 & 6.50 & 5.37 \\
0.1 & 2.0 & 7.00 & 5.02 \\
\hline
0.2 & 0.1 & 5.75 & 6.00 \\
0.2 & 0.5 & 6.05 & 5.75 \\
0.2 & 1.0 & 6.37 & 5.50 \\
0.2 & 2.0 & 6.77 & 5.23 \\
\hline
0.3 & 0.1 & 5.58 & 6.19 \\
0.3 & 0.5 & 5.89 & 5.90 \\
0.3 & 1.0 & 6.23 & 5.63 \\
0.3 & 2.0 & 6.89 & 5.12 \\
\hline
0.4 & 0.1 & 5.39 & 6.40 \\
0.4 & 0.5 & 5.74 & 6.08 \\
0.4 & 1.0 & 6.09 & 5.77 \\
0.4 & 2.0 & 6.65 & 5.33 \\
\hline
0.5 & 0.1 & 5.20 & 6.64 \\
0.5 & 0.5 & 5.58 & 6.27 \\
0.5 & 1.0 & 5.95 & 5.93 \\
0.5 & 2.0 & 6.52 & 5.45 \\
\hline
\end{tabular}
\end{center}
\caption{The marginally stable orbit and the efficiency for slowly
rotating gravastars with different spin parameters and quadrupole
moments.} \label{Tab:eff_gs}
\end{table}

These values demonstrate the variation in the location of the
inner disk edge with the changing spin parameter and quadrupole
momentum, as we have seen in the discussion on the radial
distribution of the flux. The higher these values are, the closer
the marginally stable orbits are to the center in the
dimensionless radial scale. For slowly rotating black holes, the
conversion efficiency is still close to 6\%, which is the value
obtained for Schwarzschild black holes. Up to $a_*=0.5$ it
increases to 8\%, which is still much lower than the one for
extreme Kerr black holes. For gravastars, $\epsilon$ has a smaller
variation, and always remains smaller than the conversion
efficiency for black holes. For very slow rotation, the efficiency
is approximately 5.8\%, which is close to the one for the static
black hole, and it decreases to 5\% as the quadrupole moment
increases to $2M^3$. For a higher spin parameter ($a_*\sim0.5$),
$\epsilon$ is still around 6.5\% but it becomes smaller than 6\%
as $Q$ increases. We conclude that the conversation efficiency is
higher for more rapidly rotating gravastars, but this is moderated
by the increment in its quadrupole moment. In addition to this, it
is always smaller than the $\epsilon$ for black holes, i.e.,
gravastars provide a less efficient mechanism for converting mass
to radiation than black holes.

In order that our proposal for discriminating gravastars from
black holes by using the electromagnetic emission properties of
accretion disks could be effectively applied in concrete
observational cases, it is necessary to know at least the value of
the mass and of the spin parameter of the central rotating compact
object. To estimate the spins of stellar-mass black holes in X-ray
binaries, one has to fit the continuum X-ray spectrum of the
radiation from the accretion disk, by using the standard thin disk
model, and extract the dimensionless spin parameter $a* = a/M$ of
the black hole as a parameter of the fit \cite{ShRa08}. Recently,
a number of precise black hole spin determinations have been
reported. By using Chandra and Gemini-North observations of the
eclipsing X-ray binary M33 X-7, precise values of the mass of its
black hole primary and of the system's orbital inclination angle
have been obtained. The distance to the binary is also known to a
few percent. By using these precise results, from the analysis of
15 Chandra and XMM-Newton X-ray spectra, and a fully relativistic
accretion disk model, one can find that the dimensionless spin
parameter of the black hole primary is $a*=0.77\pm0.05$
\cite{obs1}. Therefore, even that presently there are severe
observational limitations for the application of our proposal,
with the future improvements of the observational techniques, the
observation of the emission spectra of accretion disks could be
effectively used to discriminate between gravastars and black
holes.

\section{Discussions and final remarks}
\label{sec:concl}

If gravastars are surrounded by a thin shell of matter, the
presence of a turning point for matter (the point where the motion
of the infalling matter suddenly stops) at the surface of the
gravastars may have important astrophysical and observational
implications. Since the velocity of the matter at the gravastar
surface is zero, matter can be captured and deposited on the
surface of gravastar.Therefore gravastars may have a gaseous
surface, formed from a a thin layer of dense and hot gas.
Moreover, because matter is accreted continuously, the increase in
the size and density of the surface will ignite some thermonuclear
reactions \cite{YuNaRe04}. The ignited reactions are usually
unstable, causing the accreted layer of gas to burn explosively
within a very short period of time. After the nuclear fuel is
consumed, the gravastar reverts to its accretion phase, until the
next thermonuclear instability is triggered. Thus, gravastars may
undergo a semi-regular series of explosions, called type I
thermonuclear bursts, discovered first for X-ray binaries
\cite{Gr76,Tou03}.

The observational signatures indicating the presence of X-ray
bursts from gravastars are similar to those of standard neutron
stars, and are the gravitational redshift of a surface atomic
line, the touchdown luminosity of a radius-expansion burst, and
the apparent surface area during the cooling phases of the burst
\cite{Ps07}.

If the thermal radiation with wavelength $\lambda _{e}$ emitted by
the matter at the surface of the gravastar has absorption or
emission features characteristic of atomic transitions, these
features will be detected at infinity at a wavelength $\lambda
_{o}$, gravitationally redshifted with a value
\begin{equation}
z_{grav}=\frac{\lambda _{o}-\lambda _{e}}{\lambda _{e}}=e^{-\nu
\left( R\right) /2}-1,
\end{equation}
where we have assumed, for simplicity, that the gravastar is
static, and that the exterior metric can be approximated by the
standard Schwarzschild metric, given by Eq.~(\ref{Sch-metric}). By
assuming a gravastar of mass $M=4\times10^6M_{\odot}$ and radius
$R=1.4\times10^{12}$ cm, we obtain a surface redshift of
$z_{grav}=1.55$. The corresponding value of the redshift for a
neutron star with mass $M=2M_{\odot}$ and radius $R=10^6$ cm is
$z_{NS}=0.56$.  Therefore the radiation coming from the surface of
a gravastar may be highly redshifted (in standard general
relativity the redshift obeys the constrain $z\leq 2$).

Type I X -ray bursts show strong spectroscopic evidence for a
rapid expansion of the radius of the X-ray photosphere. The
luminosities of these bursts reach the Eddington critical
luminosity at which the outward radiation force balances gravity,
causing the expansion layers of the star. The touchdown luminosity
of radius-expansion bursts from a given source remain constant
between bursts to within a few percent, giving empirical
verification to the theoretical expectation that the emerging
luminosity is approximately equal to the Eddington critical
luminosity. The Eddington luminosity at infinity of a gravastar is
given by \cite{Ps07}
\begin{equation}
L_{E}^{\infty }=\left.\frac{4\pi m_{0}R^{2}}{\sigma }e^{-\lambda
}\frac{ de^{\nu /2}}{dr}\right| _{r=R},
\end{equation}
where $m_0$ is the mass of the particle and $\sigma $ is the
interaction cross section.  For the gravastar we obtain
\begin{equation}
L_{E}^{\infty }=\left.\frac{4\pi m_{0}M}{\sigma }\sqrt{1-\frac{2M}{r}}%
\right| _{r=R}.
\end{equation}

The ratio of the Eddington luminosities at infinity for a
gravastar with mass of $4\times10^6M_{\odot}$ and radius
$1.4\times10^{12}$ cm and a neutron star with mass of 2 solar
masses and radius of 10 km is $L_{Egrav}^{\infty }/L_{ENS}^{\infty
}=1.22\times 10^6$. Finally, we consider the apparent surface area
during burst cooling. Observations of the cooling tails of
multiple type I bursts from a single source have shown that the
apparent surface area of the emitting region, defined as
$S^{\infty }=4\pi D^{2}F_{c,\infty }/\sigma _{SB}T_{c,\infty }^{4}
$, where $F_{c,\infty }$ is the measured flux of the source during
the cooling tail of the burst, $T_{c,\infty }$ is the measured
color temperature of the burst spectrum, $D$ is the distance to
the source and $\sigma _{SB}$ is the Stefan-Boltzmann constant,
remains approximately constant during each burst, and between
bursts from the same source. The color temperature on the surface
of the compact object $T_{c,h}$ is related to the color
temperature measured at infinity by $T_{c,h}=T_{c,\infty }e^{-\nu
\left( R\right) /2} $ \cite{Ps07}. By introducing the color
correction factor $f_{c}=T_{c}/T_{eff}$, where $T_{eff}$ is the
effective temperature at the surface, we obtain
\begin{equation}
S^{\infty }=4\pi \frac{R^{2}}{f_{c}^{4}}\left[ z(R)+1\right] ^{2}.
\end{equation}

Since the radius of the gravastar as well as its surface redshift
may be very large quantities, the apparent area of the emitting
region as measured at infinity may be also very large. Hence all
the astrophysical quantities related to the observable properties
of the X-ray bursts, originating at the surface of the gravastar
can be calculated, and have finite values on the surface of the
gravastar and at infinity.

It was argued in \cite{YuNaRe04} that any neutron star, composed
by matter described by a more or less general equation of state,
should experience thermonuclear type I bursts at appropriate mass
accretion rates. The question asked in \cite{YuNaRe04} is whether
an ``abnormal'' surface may allow such a behavior. The gravastars
may have such a zero velocity, particle trapping, abnormal
surface. The presence of a material surface located at the ``event
horizon'' implies that energy can be radiated, once matter
collides with that surface. Thus, gravastar models, characterized
by high mass, normal matter crusts/surfaces and type I
thermonuclear bursts can be theoretically constructed. Moreover,
some of the so-called SXT's (soft X-ray transients), having a
relatively low mass function (e.g. SXT A0620-00, with a mass
function $f(M)\ge 3M_{\odot }$ \cite{YuNaRe04}), but still
exceeding the equilibrium limit of $3M_{\odot }$, or very massive
neutron stars showing the presence of a crust, may in fact be
gravastars.

It is generally expected that most of the astrophysical objects
grow substantially in mass via accretion. Recent observations
suggest that around most of the active galactic nuclei (AGN's) or
black hole candidates there exist gas clouds surrounding the
central far object, and an associated accretion disk, on a variety
of scales from a tenth of a parsec to a few hundred parsecs
\cite{UrPa95}. These clouds are assumed to form a geometrically
and optically thick torus (or warped disk), which absorbs most of
the ultraviolet radiation and the soft x-rays. The gas exists in
either the molecular or the atomic phase. Evidence for the
existence of super massive black holes comes from the very long
baseline interferometry (VLBI) imaging of molecular ${\rm H_2O}$
masers in active galaxies, like  NGC 4258 \cite{Miyo95}, and from
the astrometric and radial velocity measurements of the fully
unconstrained Keplerian orbits for short period stars around the
supermassive black hole at the center of our galaxy
\cite{Gh08,Gill09}. The VLBI imaging, produced by Doppler shift
measurements assuming Keplerian motion of the masering source, has
allowed a quite accurate estimation of the central mass, which has
been found to be a $3.6\times 10^7M_{\odot }$ super massive dark
object, within $0.13$ parsecs. Hence, important astrophysical
information can be obtained from the observation of the motion of
the gas streams in the gravitational field of compact objects.

Therefore the study of the accretion processes by compact objects
is a powerful indicator of their physical nature. However, up to
now, the observational results have confirmed the predictions of
general relativity mainly in a qualitative way. With the present
observational precision one cannot distinguish between the
different classes of compact/exotic objects that appear in the
theoretical framework of general relativity \cite{YuNaRe04}.
However, with important technological developments one may allow
to image black holes and other compact objects directly
\cite{Fa00}. Recent observations at a wavelength of 1.3 mm have
set a size of microarcseconds on the intrinsic diameter of SgrA*
\cite{Do08}. This is less than the expected apparent size of the
event horizon of the presumed black hole, thus suggesting that the
bulk of SgrA* emission may not be centered on the black hole, but
arises in the surrounding accretion flow.
A model in which Sgr A* is a compact object with a thermally
emitting surface was considered in \cite{BrNa06}. Given the very
low quiescent luminosity of Sgr A* in the near-infrared, the
existence of a hard surface, even in the limit in which the radius
approaches the horizon, places a severe constraint on the steady
mass accretion rate onto the source: $\dot{M}\leq 10^{-12}
M_{\odot}/{\rm yr}$. This limit is well below the minimum
accretion rate needed to power the observed submillimeter
luminosity of Sgr A*: $\dot{M}>10^{-10} M_{\odot}/{\rm yr}$. Thus
it follows that Sgr A* does not have a surface, i.e., that it must
have an event horizon. This argument could be made more
restrictive by an order of magnitude with microarcsecond
resolution imaging, e.g., with submillimeter very long baseline
interferometry. Submilliarcsecond astrometry and imaging of the
black hole Sgr A* at the Galactic Centre may become possible in
the near future at infrared and submillimetre wavelengths
\cite{BrLo06}. The expected images and light curves, including
polarization, associated with a compact emission region orbiting
the central black hole were computed in \cite{BrLo05}. From spot
images and light curves of the observed flux and polarization it
is possible to extract the black hole mass and spin. At radio
wavelengths, disc opacity produces significant departures from the
infrared behavior, but there are still generic signatures of the
black hole properties. Detailed comparison of these results with
future data can be used to test general relativity, and to improve
existing models for the accretion flow in the immediate vicinity
of the black hole.

With the improvement of the imaging observational techniques, it
will also be possible to provide clear observational evidence for
the existence of gravastars, and to differentiate them from other
types of compact general relativistic objects.
Indeed, in this work we have shown that the thermodynamic and
electromagnetic properties of the disks (energy flux, temperature
distribution and equilibrium radiation spectrum) are different for
these two classes of compact objects, consequently giving clear
observational signatures. More specifically, comparing the energy
flux emerging from the surface of the thin accretion disk around
black holes and gravastars of similar masses, it was found that
its maximal value is systematically lower for gravastars,
independently of the values of the spin parameter or the
quadrupole momentum. These effects are confirmed from the analysis
of the disk temperatures and disk spectra. In addition to this, it
is also shown that the conversion efficiency of the accreting mass
into radiation is always smaller than the conversion efficiency
for black holes, i.e., gravastars provide a less efficient
mechanism for converting mass to radiation than black holes. Thus,
these observational signatures may provide the possibility of
clearly distinguishing rotating gravastars from Kerr-type black
holes.

\section*{Acknowledgments}

We would like to thank the two anonymous referees for suggestions
and comments that helped us to significantly improve the
manuscript. The work of TH was supported by the General Research
Fund grant number HKU 701808P of the government of the Hong Kong
Special Administrative Region. ZK was supported by the Hungarian
Scientific Research Fund (OTKA) grant No. 69036.

\end{document}